\documentclass[aps,twocolumn,superscriptaddress,floatfix,preprintnumbers,showkeys]{revtex4}

\usepackage{graphics,amssymb,amsmath,epsfig,color}
\usepackage{graphicx}
\usepackage[ruled,vlined]{algorithm2e}
\usepackage{qcircuit}
\usepackage{pdfpages}
\usepackage{pgffor}
\graphicspath{{images/}{../images/}}

\DeclareMathOperator*{\argmin}{\arg\!\min}

\usepackage{atbegshi,etoolbox}
\makeatletter
\newcommand{\discardpages}[1]{
  \xdef\discard@pages{#1}
  \AtBeginShipout{
    \renewcommand*{\do}[1]{
      \ifnum\value{page}=##1\relax%
        \AtBeginShipoutDiscard
        \gdef\do####1{}
      \fi%
    }%
    \expandafter\docsvlist\expandafter{\discard@pages}
  }%
}
\newif\ifkeeppage
\newcommand{\keeppages}[1]{
  \xdef\keep@pages{#1}
  \AtBeginShipout{
    \keeppagefalse%
    \renewcommand*{\do}[1]{
      \ifnum\value{page}=##1\relax%
        \keeppagetrue
        \gdef\do####1{}
      \fi%
    }%
    \expandafter\docsvlist\expandafter{\keep@pages}
    \ifkeeppage\else\AtBeginShipoutDiscard\fi
  }%
}
\makeatother
\discardpages{13,15,17,19,21,23,25,27}

\begin{document}
\title{Boosting on the shoulders of giants in quantum device calibration}

\author{Alex Wozniakowski}
\email[]{wozn0001@e.ntu.edu.sg}
\affiliation{School of Physical and Mathematical Sciences, Nanyang Technological University}
\affiliation{Complexity Institute, Nanyang Technological University}
\author{Jayne Thompson}
\affiliation{Centre for Quantum Technologies, National University of Singapore}
\author{Mile Gu}
\email[]{mgu@quantumcomplexity.org}
\affiliation{School of Physical and Mathematical Sciences, Nanyang Technological University}
\affiliation{Complexity Institute, Nanyang Technological University}
\affiliation{Centre for Quantum Technologies, National University of Singapore}
\author{Felix Binder}
\affiliation{Institute for Quantum Optics and Quantum Information -- IQOQI Vienna, Austrian Academy of Sciences, Boltzmanngasse 3, 1090 Vienna, Austria}

\date{\today}

\begin{abstract}
Traditional machine learning applications, such as optical character recognition, arose from the inability to explicitly program a computer to perform a routine task. In this context, learning algorithms usually derive a model exclusively from the evidence present in a massive dataset. Yet in some scientific disciplines, obtaining an abundance of data is an impractical luxury, however; there is an explicit model of the domain based upon previous scientific discoveries. Here we introduce a new approach to machine learning that is able to leverage prior scientific discoveries in order to improve generalizability over a scientific model. We show its efficacy in predicting the entire energy spectrum of a Hamiltonian on a superconducting quantum device, a key task in present quantum computer calibration. Our accuracy surpasses the current state-of-the-art by over $20\%.$ Our approach thus demonstrates how artificial intelligence can be further enhanced by ``standing on the shoulders of giants.''
\end{abstract}

\maketitle

Prediction is paramount in almost every branch of science. In studying and designing learning systems, we are interested in prediction performance on examples unencountered during training. As machines learn inductively, generalizing training examples into an accurate model requires some restriction on the search space of hypotheses, based upon prior knowledge~\cite{Hume_1739,Haussler_1988, Mitchell_1991, Wolpert_1997}. The choice of hypothesis space constitutes the problem of inductive bias~\cite{Haussler_1988, Mitchell_1991, Caruana_1997, Wolpert_1997, Baxter_2000}, which is of broad significance in scientific applications.

Some scientific applications, such as quantum experiments, provide a paucity of data due to experimental cost, but compensate with an explicit model based upon previous discoveries~\cite{Brunton_2016, Roushan_2017, Butler_2018, Neill_2018, Chiaro_2019}. In prior work, this prior knowledge has been disregarded, and research has focused on entirely data-driven approaches that reproduce major scientific achievements or learn from toy data~\cite{Schmidt_2009, Carrasquilla_2017, Melnikov_2018, Koch-Janusz_2018, Torlai_2018, Wu_2019, Iten_2020, Wetzel_2020}. This leads us to ask if a machine learner can leverage prior scientific knowledge in order to outperform contemporary researchers? Particularly in scenarios with a shortage of experimental data.

\begin{figure*}
\centering
\begin{minipage}{\columnwidth}
\centering
\includegraphics[width=\columnwidth]{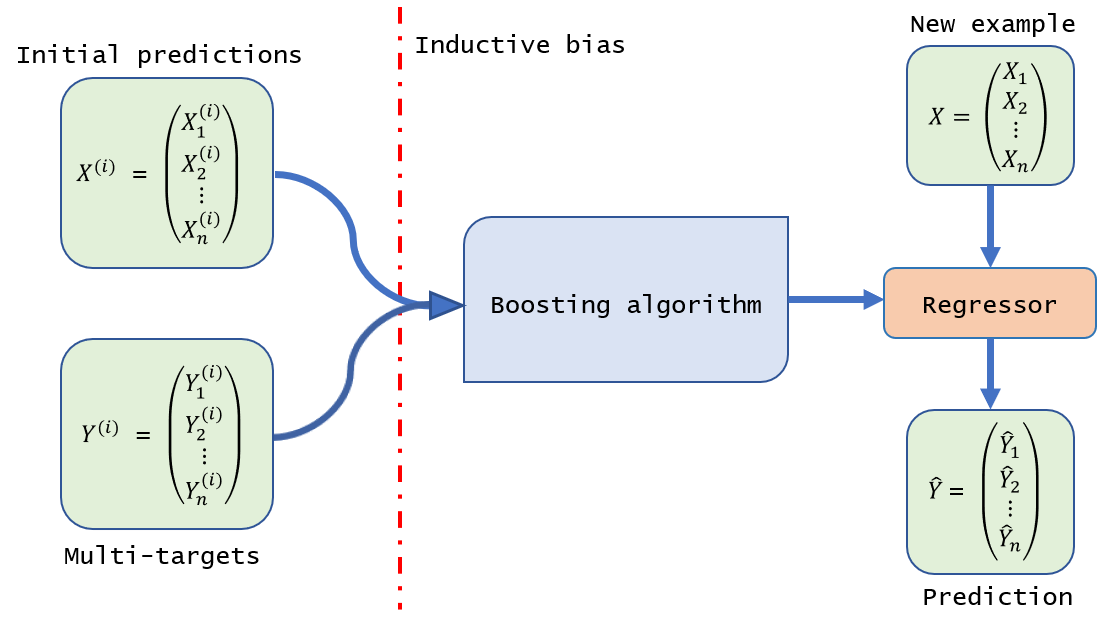}
\caption{Conceptual representation of the learning framework. Given a base regressor's initial multi-target predictions and the multi-target observations, we wrangle this data for multi-target supervised learning~\cite{Borchani_2015, Waegeman_2019}. Next, the boosting algorithm receives the training examples and acquires an inductive bias from the initial predictions. This compensates for a shortage of training examples, and the boosting algorithm improves generalizability over the base regressor. Given a new example, the boosting algorithm's returned regressor predicts a real vector.}
\label{fig:approach_diagram}
\end{minipage}\hfill
\begin{minipage}{\columnwidth}
\centering
\includegraphics[width=\columnwidth]{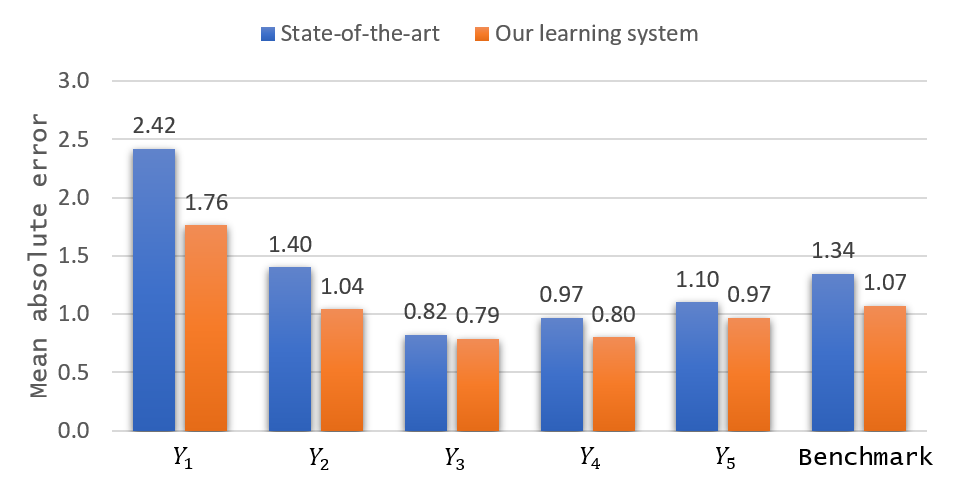} 
\caption{Benchmark task. Using the learning framework in Fig.~\ref{fig:approach_diagram}, our learning system surpasses the state-of-the-art~\cite{Roushan_2017, Neill_2018, Chiaro_2019} by over $20\%$ on the calibration task of simultaneously predicting the entire energy spectrum of a Hamiltonian Eq.~\ref{eq:bose_hubbard} on a nearest-neighbor coupled linear chain of superconducting qubits. Moreover, our learning system outperforms the state-of-the-art on each individual prediction task, i.e., $Y_{j},$ where $j \in \{1,2,\dots,5\}.$}
\label{fig:test_data_mae}
\end{minipage}
\end{figure*}

Here, we introduce a new framework that restricts a learning algorithm's search space of hypotheses. It does so by leveraging prior knowledge contained in predictions generated by a scientific model (see Fig.~\ref{fig:approach_diagram}). 
In contrast to conventional supervised learning, we focus on the simultaneous prediction of multiple real variables, which is known as multi-target regression~\cite{Caruana_1997, Sklearn_2011, Borchani_2015, Waegeman_2019}. This enables the learning algorithm to improve generalizability over the scientific model by discovering relationships among the targets, which the model did not envisage. In principle, this approach shares similarities with neuroplasticity, whereby the nervous system is able to adapt and optimize its limited resources in response to sensory experiences~\cite{Pascual_leone_2005}. 

To test our learning system, we establish a proxy of expert human-level performance on the calibration \textit{benchmark task} of simultaneously predicting the entire energy spectrum of a Hamiltonian on a superconducting quantum device~\cite{Chen_2014, Roushan_2017, Neill_2018, Chiaro_2019}. In this scenario, there is a shortage of data due to operational cost of the experiment~\cite{Roushan_2017}. The explicit scientific model of the device's quantum behavior is state-of-the-art \cite{Roushan_2017, Neill_2018, Chiaro_2019}. We demonstrate that our learning system surpasses this baseline of expert human-level performance by over $20\%$ (see Fig.~\ref{fig:test_data_mae}). Consequently, we advance the current ability to precisely generate Hamiltonians with programmable parameters for a variety of quantum simulation applications. Our result complements other recent applications of machine learning in scientific settings, and more specifically quantum systems~\cite{Schmidt_2009, Zahedinejad_2016, Brunton_2016, Lin_2017, Biamonte_2017, Carrasquilla_2017, Koch-Janusz_2018, Torlai_2018, Butler_2018, Melnikov_2018, Dunjko_2018, Wu_2019, Giuseppe_2019, Mehta_2019, Iten_2020, Wetzel_2020, Udrescu_2020}. To interpret our results we use techniques from explainable machine learning~\cite{Lundberg_2017, Molnar_2020} to uncover parameter dependencies in the original scientific model.

\section*{Results}

\textbf{Benchmark task} -- In order to establish a proxy of expert human-level performance for the analysis of our learning system, we study a superconducting qubit architecture~\cite{Chen_2014}. The quantum device is a nearest-neighbor coupled linear chain of superconducting qubits with tunable qubit frequencies and tunable inter-qubit interactions~\cite{Chen_2014, Roushan_2017, Neill_2018, Chiaro_2019}. Each qubit is embedded in the subspace spanned by the ground state and first excited state of a nonlinear photonic resonator in the microwave regime. The total Hamiltonian of the device is approximately described by the Bose-Hubbard model truncated at two local excitations
\begin{align}
\begin{split}
\mathcal{H} & = \sum_{j=1}^{n} \delta_{j} \hat{a}^{\dagger}_{j} \hat{a}_{j} + \frac{L}{2} \hat{a}^{\dagger}_{j} \hat{a}_{j} (\hat{a}^{\dagger}_{j} \hat{a}_{j} - 1) \\
& + \sum_{j=1}^{n-1} g_{j, j+1} (\hat{a}^{\dagger}_{j} \hat{a}_{j+1} + \hat{a}_{j} \hat{a}^{\dagger}_{j+1}),
\end{split}
\label{eq:bose_hubbard}
\end{align} where $n > 1$ is the number of qubits, $\hat{a}^{\dagger}$ ($\hat{a}$) is the bosonic creation (annihilation) operator, $\delta_{j}$ is the random on-site detuning, $L$ is the on-site Hubbard interaction, and $g_{j, j+1}$ is the hopping rate between nearest neighbor lattice sites. Quantum evolution is typically realized by allowing the entire system to interact at once, which also admits translation into the prototypical quantum circuit model~\cite{Neill_2018}. 

In the benchmark task, the device contains $9$ qubits. The $n=5$ rightmost qubits and $4$ interleaving couplers were utilized during experimentation, while the $4$ leftmost qubits and couplers were left idle. The device is being calibrated for a many-body localization experiment~\cite{Roushan_2017, Chiaro_2019}, where different relaxation dynamics are observed, depending on the extent of random disorder in the system. Probing this quantum phenomenon requires study of the entire energy spectrum, which be achieved experimentally through many-body Ramsey spectroscopy~\cite{Roushan_2017}. 

Here, we focus on the identification of $5$ eigenenergies belonging to Eq.~\ref{eq:bose_hubbard}, when it describes hopping of a single photon in a disordered potential. The energy eigenstates are generally not local and each instance of the many-body Ramsey spectroscopy technique sorts the measured eigenenergies in ascending order:\ $Y_{1}, Y_{2},\dots,Y_{5}.$ In the present context of machine learning, we refer to these variables as single-targets and to their collection as a multi-target (details in Methods). 

The calibration is performed in two steps~\cite{Roushan_2017, Neill_2018, Chiaro_2019}, where the benchmark dataset pertains to the second step. In the first step, the room temperature time-dependent pulses that orchestrate the computation are calibrated to arrive at the device:\ orthogonally, synchronously, and without pulse-distortion~\cite{Neill_2018}. In the second step, the control pulses are converted to matrix elements of the Hamiltonian Eq.~\ref{eq:bose_hubbard}. Underlying this conversion is a finitely parameterized model of the device's electronic circuitry, which is directly encoded in the classical control program~\cite{Roushan_2017, Neill_2018, Chiaro_2019}. 

Inferring the physical parameters of the control model entails fitting the two lowest transition energies of each qubit as a function of qubit and coupler flux-biases~\cite{Neill_2018}. Next, the many-body Ramsey spectroscopy technique benchmarks the collective dynamics of the device, where all of the qubits are coupled and near resonance with each other~\cite{Roushan_2017, Chiaro_2019}. Then, minimization of the absolute error loss function, which compares the multi-targets with the multi-target predictions generated by the classical control program, numerically optimizes the physical parameters (see Eq.~\ref{eq:absolute_error_loss_function} in Methods). Lastly, the updated classical control program generates $136$ multi-target predictions for the $136$ multi-targets in the benchmark task. Using $41$ of these multi-targets and the corresponding predictions, we compute the mean absolute errors for the single-targets and the average mean absolute error for the multi-targets (see Eq.~\ref{eq:single_target_test_error} and Eq.~\ref{eq:multi_target_test_error} in Methods). We refer to the $1.34$ MHz average mean absolute error as the benchmark error in Fig.~\ref{fig:test_data_mae}. Using this benchmark error, we establish a proxy of expert human-level performance on the benchmark task. As the estimated optimal error rate, set by the coherence time of the device, is $1$MHz~\cite{Roushan_2017}, we ask the algorithm design question:\ can we do better? 

Using the classical control program, this would require us to directly write the higher order terms in the Hamiltonian, environmental interactions, manufacturing or operational errors, etc., for every recalibration. Clearly this strategy is impractical within recalibration timescales~\cite{Linke_2017, Kelly_2018_1}. Therefore, we propose a paradigm shift, whereby we incorporate the prior knowledge in the classical control program into a boosting algorithm whose primary goal is to discover a more accurate model of the domain (see Fig.~\ref{fig:test_data_mae}). In this way, we can feedback improved multi-target predictions to the optimization step in the calibration process and update the physical parameters in the control model~\cite{Neill_2018}. Thus, enhancing the ability to generate Hamiltonians with programmable parameters for a variety of quantum simulation applications.

\textbf{Learning Framework} -- Multi-target regression aims to simultaneously predict multiple real variables, and research in this direction is intensifying~\cite{Borchani_2015, Waegeman_2019}. Here, we introduce a two-step stacking~\cite{Wolpert_1992, Breiman_1996, Breiman_1997_2, Borchani_2015, Waegeman_2019} framework that supplies a boosting algorithm with an inductive bias contained in the initial multi-target predictions generated by a base regressor (details in Methods). In essence, the base regressor acts as data preprocessor and the boosting algorithm assays to improve generalization performance by discovering relationships among the single-targets. This approach is related to multi-target regularization, which reduces the problem of overfitting~\cite{Breiman_1997_2, Borchani_2015, Waegeman_2019}, as well as methods in deep learning, such as pre-training~\cite{Erhan_2010} and weight sharing~\cite{Caruana_1997}. 

\begin{figure}
\centering
\includegraphics[width=\columnwidth]{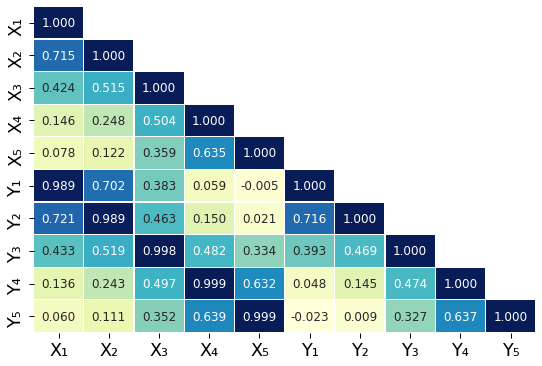}
\caption{Pairwise correlations in the training examples. We denote the features in each example by $X_{j},$ and the single-targets in each multi-target by $Y_{j},$ where $j \in \{1,2,\dots,5\}$ is the number of superconducting qubits utilized in the benchmark task~\cite{Roushan_2017, Neill_2018, Chiaro_2019}.}
\label{fig:heatmap}
\end{figure}

In applying the learning framework to the benchmark dataset, the first step wrangles the data for multi-target supervised learning~\cite{Borchani_2015, Waegeman_2019}. Namely, we regard a multi-target prediction generated by the classical control program~\cite{Roushan_2017, Neill_2018, Chiaro_2019} as an example and the associated instance of the many-body Ramsey spectroscopy technique~\cite{Roushan_2017} as the label. Under the distribution-free setting~\cite{Haussler_1992, Kearns_1994_1, Kearns_1994_2, Friedman_2001, Friedman_2003, Hastie_2009}, we split the labeled examples into $m_{train}=95$ and $m_{test}=41$ ordered pairs for training and test data, respectively, where the choice of splitting fraction is a heuristic~\cite{Ng_2020, Hastie_2009}. In the second step, a boosting algorithm receives the training examples with pairwise correlations shown in Fig.~\ref{fig:heatmap}, and we request a multi-target regressor $\hat{h}$ as output. 
The boosting algorithm proceeds by reducing the multi-target regression task to $5$ independent single-target regression subtasks~\cite{Borchani_2015, Waegeman_2019}. For the $jth$ single-target regression subtask, the $j$th single-target boosting algorithm induces the single-target regressor $\hat{h}_{j}$ on the $j$th slice of the training examples, where $j=\{1,2,\dots,5\}$ (see Eq.~\ref{eq:transformed_training_data} in Methods). Subsequently, the boosting algorithm concatenates the single-target regressors into a multi-target regressor. Given a new example $X,$ the multi-target regressor predicts a $5$-dimensional real vector $\hat{Y} = \hat{h}(X).$ 

\textbf{Gradient boosting prior knowledge} -- Boosting is an algorithmic paradigm for improving the performance of any given learning algorithm, interconnecting machine learning~\cite{Powell_1987, Kearns_1988, Hastie_1990, Schapire_1990, Kearns_1994_3, Freund_1995, Freund_1997, Breiman_1997_1, Mason_1999, Schapire_2002, Chen_2016, Ke_2017}, statistics~\cite{Tukey_1977, Friedman_2000, Friedman_2001, Friedman_2003, Hastie_2009} and signal processing~\cite{Mallat_1993, Vincent_2002, Donoho_2012} through the study of additive expansions~\cite{Hastie_1990, Friedman_2000, Friedman_2001, Hastie_2009}. Gradient boosting is a generic version of boosting, which is widely used in practice~\cite{Friedman_2001, Friedman_2003, Hastie_2009, He_2014, Chen_2016, Ke_2017}, and the additive expansion is designed to finesse the curse of dimensionality and provide flexibility over linear models~\cite{Hastie_1990, Friedman_2000, Friedman_2001, Friedman_2003, Hastie_2009}. Nonetheless, the standard form of gradient boosting does not allow for the direct incorporation of prior knowledge, which is essential in the benchmark task.

Here, we propose a modification of the standard additive expansion~\cite{Friedman_2001, Friedman_2003, Hastie_2009, He_2014, Chen_2016, Ke_2017} for the $j$th single-target regression subtask 
\begin{equation}
\label{eq:additive_model}
h_{j}(X;\{\alpha_{j}, \theta_{j}\}) = X_{j} + \sum_{k=1}^{K_{j}} \alpha_{j,k} b (X; \theta_{j,k})),
\end{equation}
where the collection of expansion coefficients $\alpha_{j,k}$ and parameter sets $\theta_{j,k}$ is given by
$\{\alpha_{j}, \theta_{j}\} = \{\alpha_{j,1},\dots, \alpha_{j,K_{j}},\theta_{j,1},\dots,\theta_{j,K_{j}}\},$
and $K_{j}$ denotes the number of real-valued basis functions $b(X;\theta_{j,k})$ of the example $X$ (details in Methods). In the standard additive expansion, the first term is a constant offset value that does not depend upon the example, and it is usually determined by maximum likelihood estimation~\cite{Hastie_1990, Friedman_2000, Friedman_2001, Friedman_2003, Hastie_2009}. In the work of Schapire et al, prior knowledge was incorporated into the G\"{o}del prize winning AdaBoost algorithm by modifying the loss function for single-target classification tasks~\cite{Schapire_2002}. In machine learning, the basis function is called a weak learner~\cite{Kearns_1988, Schapire_1990, Kearns_1994_3, Freund_1995, Breiman_1997_1, Freund_1997, Schapire_2002}, and the predominant choice is a shallow decision tree~\cite{Friedman_2001, Friedman_2003, Hastie_2009, He_2014, Chen_2016, Ke_2017}. Taking a reroughing viewpoint~\cite{Tukey_1977}, Eq.~\ref{eq:additive_model} decomposes the $j$th single-target into a smooth term, i.e., the first term, and a noise term, i.e., the linear sum of basis functions. In the application, the classical control program~\cite{Roushan_2017, Neill_2018, Chiaro_2019} generates the smooth term and the noise term adaptively models the relationships between the single-targets in Fig.~\ref{fig:heatmap} without overwhelming the prior knowledge (see Supplementary Information).

In practice, fitting an additive expansion by minimizing the data-based estimate of the $j$th single-target expected loss is usually infeasible~\cite{Mallat_1993, Friedman_2000, Friedman_2001, Vincent_2002, Friedman_2003, Hastie_2009, Donoho_2012} (see Eq.~\ref{eq:single_target_expected_loss} in Methods). Here, we employ a greedy stagewise algorithm to approximate this optimization problem, whereby the stagewise algorithm sequentially appends basis functions to the additive expansion without adjusting the previously learned expansion coefficients or parameter sets, as opposed to a stepwise algorithm~\cite{Mallat_1993, Friedman_2000, Friedman_2001, Vincent_2002, Friedman_2003, Hastie_2009, Donoho_2012} (see Alg.~\ref{alg:1} in Methods). As a result of modifying the standard additive expansion in Eq.~\ref{eq:additive_model}, the learning framework directly incorporates prior knowledge into gradient boosting~\cite{Friedman_2001, Friedman_2003, Hastie_2009, He_2014, Chen_2016, Ke_2017} by changing the initialization step (details in Methods). As an aside, this idea can be applied in compressed sensing by similarly changing the initialization step in matching pursuit and its extensions~\cite{Mallat_1993, Vincent_2002, Donoho_2012}. 

\textbf{Inbuilt model selection} -- The greedy stagewise algorithm does not always improve performance over the smooth term. Hence, we introduce an augmented version with inbuilt model selection, which scores the incumbent smooth term and the candidate greedy stagewise algorithm with a modification of $k$-fold cross-validation (details in Methods). If the incumbent performs better or equally well, then the augmented version returns the smooth term as the induced single-target regressor. Otherwise, the augmented version calls the candidate (see Alg.~\ref{alg:2} in Methods).

\begin{figure}
\centering
\includegraphics[width=0.9\columnwidth]{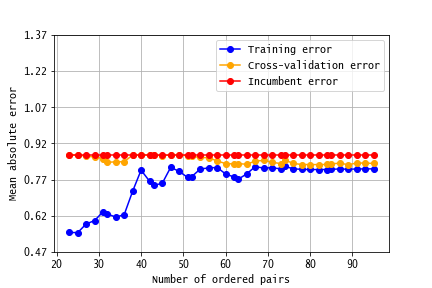}
\caption{Augmented learning curve for the single-target $Y_{3}.$ We show the training, cross-validation, and incumbent errors for varying amounts of training examples in blue, orange, and red, respectively. As a consequence of the inbuilt model selection step, the cross-validation error is bounded from above by $0.87$ MHz. With $51,$ or more, ordered pairs the candidate greedy stagewise algorithm always outperforms the incumbent smooth term~\cite{Roushan_2017, Neill_2018, Chiaro_2019}.}
\label{fig:eig_learning_curve}
\end{figure}

In Fig.~\ref{fig:eig_learning_curve}, we illustrate the model selection step with an augmented learning curve for the single-target regression subtask $Y_{3}$ with training sizes varying between $23$ to $95$ ordered pairs.
Here, the augmented learning curve shows the incumbent error (red) in addition to the training and cross-validation errors (blue and yellow) shown in a prototypical learning curve~\cite{Hastie_2009, Sklearn_2011, Ng_2020}. The incumbent error bounds the cross-validation error from above. As the training size increases, the training error tends to increase, the cross-validation error tends to decrease, and both errors exhibit random fluctuations, which typically occur with less than $100$ ordered pairs~\cite{Ng_2020}. When there are less than $51$ ordered pairs, the incumbent usually performs better, whereas the candidate always outperforms the incumbent with $51,$ or more, ordered pairs. 

For the single-target regression subtasks $Y_{1},$ $Y_{2},$ and $Y_{5},$ the candidate always performs better, and in general the candidate always performs better with $60,$ or more, ordered pairs (see Supplementary Information). Thus, the boosting algorithm used the greedy stagewise algorithm in each single-target regression subtask in Fig.~\ref{fig:test_data_mae}, where the boosting algorithm outperforms the baseline of expert human-level performance~\cite{Roushan_2017, Neill_2018, Chiaro_2019} by over $20\%.$

\textbf{Examining the prior knowledge} -- Data preprocessing can significantly impact generalization performance, especially if there is a shortage of training examples~\cite{Erhan_2010, Sklearn_2011}. Here, we examine the classical control program~\cite{Roushan_2017, Neill_2018, Chiaro_2019} as a data preprocessor for the downstream boosting algorithm, whereby the classical control program transforms $5$ qubit and $4$ coupler bias features from an instance of the spectroscopy protocol~\cite{Roushan_2017} into an initial multi-target prediction (see Supplementary Information). Namely, we regard a collection of $5$ qubit and $4$ coupler bias features as an example, and we induce a fully-connected neural network~\cite{Sklearn_2011} for each single-target. Next, we apply the SHAP framework to approximate each induced neural network with a simpler linear explanation model~\cite{Lundberg_2017} (see Eq.~\ref{eq:explanation_model} in Methods). The linear coefficients, known as SHAP values, allocate the importance of each feature for each single-target training data prediction~\cite{Lundberg_2017, Molnar_2020}. 

\begin{figure}
\centering
\includegraphics[width=0.9\columnwidth]{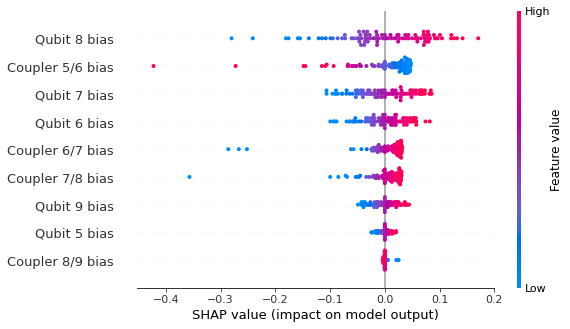}
\caption{SHAP summary plot~\cite{Lundberg_2017} for the single-target $Y_{1}.$ Using the collection of $9$ qubit and coupler bias features as an example, we induce a fully connected neural network~\cite{Sklearn_2011} to predict $Y_{1}.$ The horizontal axis is centered at the average training example prediction, and the vertical axis ascendingly orders the features according to their importance. Each point is a SHAP value for a particular example, the coloring represents the bias value, and overlapping points are randomly jittered along the vertical axis to avoid collisions~\cite{Lundberg_2017, Molnar_2020}.}
\label{fig:shap}
\end{figure}

In Fig.~\ref{fig:shap}, we acquire an overview of each feature's importance and effect in the single-target regression subtask $Y_{1}$~\cite{Lundberg_2017, Molnar_2020} (see Eq.~\ref{eq:feature_importance} in Methods; see Supplementary Information for additional SHAP summary plots).
The features are ascendingly ordered from bottom to top according to their importance, a point represents a SHAP value, and the coloring represents the bias value, e.g., a reddish point for the coupler $5/6$ bias feature illustrates strong coupling at the coupler between the $5$th and $6$th qubit sites. As can be clearly seen in Fig.~\ref{fig:shap}, the qubit $8$ bias is the most important feature, which corresponds to an interior qubit site near the physical boundary of the linear chain. The coupler $5/6$ bias is the only coupler bias in the top $4$ features. 

In comparison, the coupler $5/6$ bias is the most important feature in the single-target regression subtasks $Y_{3}$ and $Y_{4},$ and the coupler $8/9$ bias is the most important feature in the single-target regression subtasks $Y_{2}$ and $Y_{5}$ (see Supplementary Information). The former feature corresponds to a coupler near the experimentally imposed boundary of the linear chain, and the latter feature corresponds to a coupler near the physical boundary of the linear chain. Whereas the qubit biases, which correspond to interior qubit sites, are $3$ out of the $4$ most important features in the single-target regression subtask $Y_{1},$ the only other single-target regression subtask with a qubit bias in the top $4$ features is $Y_{2}.$ 

This feature dependence merits some discussion. As each instance of the spectroscopy protocol~\cite{Roushan_2017} ascendingly orders the eigenenergies, one might expect that on average over all runs the feature dependence would be qualitatively the same for each single-target. Indeed, under independent and identically distributed sampling of the input parameters we would expect the data to exhibit a symmetry under permutation among the local bias and coupling parameters in Eq.~\ref{eq:bose_hubbard}. In line with this intuition, we observe a noticeably marked dependence on the coupler bias features closest to the physical boundaries for all single-targets. However, more generally, the permutation symmetry is broken in the benchmark dataset, not least because the model consists of few sites and is patently not well approximated by closed boundary conditions. Some of the individual single-targets, for instance, have a stronger dependence on specific on-site biases than others. This suggest that different sites correlate more strongly with larger or smaller eigenenergies. An example is the aforementioned strong dependence of the most negative eigenenergy $Y1$ on the on-site bias at site $8$. We attribute this to the geometry of the physical configuration and note that this asymmetric feature dependence is already present in the initial multi-target predictions generated by the data preprocessor.

\section*{Discussion}

While entirely data-driven approaches are successful in machine learning applications with an abundance of data, these machine learning methods break down in scenarios with a shortage of data. Overcoming this obstacle requires some resource that compensates for the lack of data~\cite{Schapire_2002}. In quantum device calibration applications, data accumulation is low~\cite{Roushan_2017}, but there is an analytical model of the domain based upon prior scientific discoveries. Our result demonstrates that a machine learner can refine and enhance such discoveries with a minuscule amount of real experimental data. Using this approach, our learning system surpassed its scientific contemporaries~\cite{Roushan_2017, Neill_2018, Chiaro_2019} by over $20\%$ on the superconducting quantum device calibration task, thereby providing a pathway for the successful interface of artificial intelligence and physics. Moreover, we have demonstrated the robustness of our approach by incorporating inbuilt model selection and we have established a diagnostic method to examine the underlying scientific model with SHAP learning techniques~\cite{Lundberg_2017, Molnar_2020}.

Although we have focused on a quantum device calibration application, the presented machine learning approach can have significant impact further afield. We have introduced an additive expansion in Eq.~\ref{eq:additive_model} that is a modification of a model at the heart of several function approximation methods in engineering~\cite{Powell_1987}, machine learning~\cite{Powell_1987, Schapire_1990, Freund_1995, Freund_1997, Breiman_1997_1, Mason_1999, Chen_2016, Ke_2017}, statistics~\cite{Tukey_1977, Hastie_1990, Friedman_2000, Friedman_2001, Friedman_2003, Hastie_2009} and signal processing~\cite{Mallat_1993, Vincent_2002, Donoho_2012}. Gradient boosting is one of the most popular learning algorithms in data science and machine learning competitions~\cite{Chen_2016, Ke_2017}, and also in real-world production pipelines~\cite{He_2014}. Our approach enables it to take advantage of prior knowledge, especially when data is scarce. Other potential applications include compressed sensing, where prior knowledge about sparsity has resulted in an advantage over the Nyquist-Shannon sampling theorem~\cite{Mallat_1993, Vincent_2002, Donoho_2012}. Indeed, physical manifestations of Occam's razor, symmetry and complexity have already significantly influenced the development of learning and prediction~\cite{Shalizi_2001, Gu_2012, Lin_2017, Udrescu_2020} -- and thus a systematic approach to incorporating prior scientific knowledge into a machine learner provides a natural advancement of the mutualistic relationship between human researchers and artificial intelligence.

\section*{Acknowledgements}
We are grateful to Benjamin Chiaro, who ran the experiment, collected the data, and shared it with us during his time as a graduate student at UC Santa Barbara, and to Pedram Roushan for helpful discussions. This work is supported by the Singapore Ministry of Education Tier $1$ grant RG$162$/$19$, Singapore National Research Foundation Fellowship NRF-NRFF$2016$-$02$ and NRF-ANR grant NRF$2017$-NRF-ANR$004$ VanQuTe, and the FQXi large grants:\ the role of quantum effects in simplifying adaptive agents and are quantum agents more energetically efficient at making predictions? A.W.\ was partially supported by the Grant TRT $0159$ on mathematical picture language from the Templeton Religion Trust and thanks the Academy of Mathematics and Systems Science (AMSS) of the Chinese Academy of Sciences for their hospitality, where part of this work was done. F.C.B.\ acknowledges funding from the European Union’s Horizon $2020$ research and innovation programme under the Marie Skłodowska-Curie Grant Agreement No.\ $801110$ and the Austrian Federal Ministry of Education, Science and Research (BMBWF).

\section*{Methods}

\textbf{Multi-target regression background} -- In the setting of our learning framework, let $\mathcal{X}$ be the domain, where we refer to points in $\mathcal{X}$ as examples. Let $\mathcal{Y} \subseteq \mathbb{R}^{n}$ be the target space of multi-target observations, where we refer to vectors in $\mathcal{Y}$ as multi-targets and to components of vectors as single-targets. We refer to an ordered pair in the product of the domain and the target space $(X,Y) \in \mathcal{X} \times \mathcal{Y}$ as a labeled example. Moreover, we are given a finite sequence of labeled examples
\begin{equation}
\label{eq:labeled_examples}
S = \{ (X^{(i)}, Y^{(i)}) \}_{i=1}^{m} \in (\mathcal{X} \times \mathcal{Y})^{m},
\end{equation}
which is supposed random so that there is an unknown probability distribution on $\mathcal{X} \times \mathcal{Y}$~\cite{Haussler_1992, Kearns_1994_1, Kearns_1994_2}.  

We wish to find some simple pattern in the labeled examples, namely a multi-target regressor 
$ h: \mathcal{X} \to \mathcal{Y}.$
However, there may be no functional relationship between the domain and the target space in this agnostic setting~\cite{Kearns_1994_1, Kearns_1994_2}. In order to measure the predictive prowess of a multi-target regressor, we introduce the decision theoretic concept of a loss function~\cite{Haussler_1992, Kearns_1994_1, Kearns_1994_2, Friedman_2001, Friedman_2003, Hastie_2009}, where we denote a non-negative multi-target loss function by 
$\ell: \mathcal{Y} \times \mathcal{Y} \to \mathbb{R}_{\geq 0}.$
Given a labeled example $(X,Y) \in \mathcal{X} \times \mathcal{Y},$ the loss of some multi-target regressor $h$ on the labeled example is denoted by $\ell(Y, h(X)).$ The multi-target loss function measures the magnitude of error in predicting $h(X),$ when the multi-target is $Y.$ 

Here, we study loss functions that are decomposable over the targets, which provides a joint target view~\cite{Borchani_2015, Waegeman_2019}. Let $\mathcal{Y}_{j} \subseteq \mathbb{R}$ be the single-target space of the $j$th single-target observations. We denote a single-target regressor by
$ h_{j}: \mathcal{X} \to \mathcal{Y}_{j}.$ We denote a nonnegative single-target loss function by 
$\ell_{j}: \mathcal{Y}_{j} \times \mathcal{Y}_{j} \to \mathbb{R}_{\geq 0}.$
Given a labeled example in the product of the domain and the single-target space $(X, Y_{j}) \in \mathcal{X} \times \mathcal{Y}_{j},$ the loss of some single-target regressor $h_{j}$ on the labeled example is denoted $\ell_{j}(Y_{j}, h_{j}(X)).$ The single-target loss function measures the magnitude of error in predicting $h_{j}(X),$ when the single-target is $Y_{j}.$ We define a loss function that is decomposable over the targets by  
\begin{equation}
\ell(Y, h(X)) = \sum_{j=1}^{n} \ell_{j}(Y_{j}, h_{j}(X)),
\end{equation}
in accord with~\cite{Borchani_2015, Waegeman_2019}. In the application, we study the absolute error loss function, which is decomposable over the targets. Namely,
\begin{align}
\label{eq:absolute_error_loss_function}
\begin{split}
\ell(Y, h(X)) &= ||Y - h(X) ||_{1} \\
&= \sum_{j=1}^{n} |Y_{j} - h_{j}(X)| \\
&= \sum_{j=1}^{n} \ell_{j}(Y_{j}, h_{j}(X)),
\end{split}
\end{align}
where $||\cdot||_{p}$ denotes the $\mbox{L}^{p}$ norm. Using the joint view, the multi-target regression task reduces to $n$ independent single-target regression subtasks
\begin{align}
\mathbb{E}_{X,Y} [\ell(Y, h(X))] &= \mathbb{E}_{X,Y} [\sum_{j=1}^{n} \ell_{j}(Y_{j}, h_{j}(X))] \nonumber \\
&= \sum_{j=1}^{n} \mathbb{E}_{X,Y_{j}} [\ell_{j}(Y_{j}, h_{j}(X))] \label{eq:single_target_expected_loss},
\end{align}
where the first line follows from the choice of a loss function that is decomposable over the targets, and the second line follows from linearity~\cite{Borchani_2015, Waegeman_2019}. In this case, the optimal $j$th single-target regressor is the one that minimizes the $j$th single-target expected loss
\begin{equation}
h^{*}_{j} = \argmin_{h_{j}} \mathbb{E}_{X,Y_{j}} [\ell_{j}(Y_{j}, h_{j}(X))].
\end{equation}

Under the distribution-free setting, the $j$th single-target expected loss is not available~\cite{Haussler_1992, Kearns_1994_1, Kearns_1994_2, Friedman_2001, Friedman_2003, Hastie_2009}. Consequently, we split Eq.~\ref{eq:labeled_examples} into training, validation, and test data, if there is sufficient data for an explicit validation stage. Otherwise, we forgo the validation split. Here, we focus on the case of splitting Eq.~\ref{eq:labeled_examples} into $m_{train}$ and $m_{test}$ ordered pairs for training and test data, respectively, as there is a shortage of labeled examples in the application. Moreover, we isolate the test data from the training data, whereby training data is recyclable and test data is single-use. Using the test data, we approximate the $j$th single-target expected loss with the mean absolute error 
\begin{equation}
\label{eq:single_target_test_error}
\frac{1}{m_{test}} \sum_{i=1}^{m_{test}} |Y^{(i)}_{j} - h_{j}(X^{(i)})|.
\end{equation}
Then, we approximate the expected loss with the average mean absolute error
\begin{equation}
\label{eq:multi_target_test_error}
\frac{1}{n} \sum_{j=1}^{n} \frac{1}{m_{test}} \sum_{i=1}^{m_{test}} |Y^{(i)}_{j} - h_{j}(X^{(i)})|,
\end{equation}
and we refer to this error as the benchmark error in Fig.~\ref{fig:test_data_mae}.

\textbf{Two-step stacking framework} -- In the learning framework, let $\mathcal{X} \subseteq \mathbb{R}^{n}$ be the domain of initial multi-target predictions generated by a base regressor. We assume the availability of these predictions as well as the associated multi-target observations. In this way, the learning framework can be applied in tandem with scientific models (see the Supplementary Information for a brief review of the traditional two-step stacking approach). 

In the first step, we wrangle the labeled examples Eq.~\ref{eq:labeled_examples}, and we represent them with an $m \times 2n$ design matrix
\begin{equation}
\label{eq:labeled_examples_matrix}
\begin{pmatrix}
\rule[.5ex]{3.5em}{0.4pt} X^{(1)} \rule[.5ex]{3.5em}{0.4pt} & \vline & \rule[.5ex]{3.5em}{0.4pt} Y^{(1)} \rule[.5ex]{3.5em}{0.4pt}\\  
\rule[.5ex]{3.5em}{0.4pt} X^{(2)} \rule[.5ex]{3.5em}{0.4pt} & \vline & \rule[.5ex]{3.5em}{0.4pt} Y^{(2)} \rule[.5ex]{3.5em}{0.4pt}\\
\vdots & \vline & \vdots\\
\rule[.5ex]{3.5em}{0.4pt} X^{(m)} \rule[.5ex]{3.5em}{0.4pt} & \vline & \rule[.5ex]{3.5em}{0.4pt} Y^{(m)} \rule[.5ex]{3.5em}{0.4pt}\\  
\end{pmatrix},
\end{equation}
where $m$ denotes the number of multi-targets and $n$ denotes the number of single-targets. Next, we split Eq.~\ref{eq:labeled_examples_matrix} into $m_{train}$ and $m_{test}$ rows for training and test data, respectively. In the second step, the boosting algorithm receives the training data, which has shape $m_{train} \times 2n,$ and we request a multi-target regressor as output. In the $j$th single-target regression subtask, the boosting algorithm slices the $j$th single-target from the training data
\begin{equation}
\label{eq:transformed_training_data}
\begin{pmatrix}
\rule[.5ex]{3.5em}{0.4pt} X^{(1)} \rule[.5ex]{3.5em}{0.4pt} & \vline & Y_{j}^{(1)}\\  
\rule[.5ex]{3.5em}{0.4pt} X^{(2)} \rule[.5ex]{3.5em}{0.4pt} & \vline & Y_{j}^{(2)}\\
\vdots & \vline & \vdots\\
\rule[.5ex]{2.5em}{0.4pt} X^{(m_{train})} \rule[.5ex]{2.5em}{0.4pt} & \vline & Y_{j}^{(m_{train})}\\  
\end{pmatrix},
\end{equation}
where the matrix has shape $m_{train} \times (n + 1).$ Next, the single-target boosting algorithm detailed in Alg.~\ref{alg:2} induces the $j$th single-target regressor $\hat{h}_{j}$ on Eq.~\ref{eq:transformed_training_data}. After completion of each single-target regression subtask, the boosting algorithm concatenates the induced single-target regressors into the multi-target regressor
$\hat{h} = (\hat{h}_{1}, \hat{h}_{2}, \dots, \hat{h}_{n})^{T}$. Given a new example $X \in \mathcal{X},$ the multi-target regressor predicts an $n$-dimensional real vector $\hat{Y} = \hat{h}(X)$ (see Fig.~\ref{fig:approach_diagram}).

\textbf{Model selection} -- As the test data is single-use, we need to simultaneously select the best performing single-target boosting algorithm detailed in Alg.~\ref{alg:1} for the $j$th single-target regression subtask and estimate the $j$th mean absolute error Eq.~\ref{eq:single_target_test_error}, where $j \in \{ 1,2,\dots,n \}.$ Moreover, we need to ensure that the selected $j$th single-target boosting algorithm is able to choose the smooth term, if the noise term in Eq.~\ref{eq:additive_model} degrades performance (see Fig.\ \ref{fig:eig_learning_curve}). For this objective, we review nested cross-validation~\cite{Hastie_2009, Cawley_2010, Sklearn_2011}, and we describe the modification of $k$-fold cross-validation utilized in Alg.~\ref{alg:2}, which is similar to learning algorithms with inbuilt cross-validation~\cite{Sklearn_2011}. 

\begin{figure}
\centering
\includegraphics[width=0.9\columnwidth]{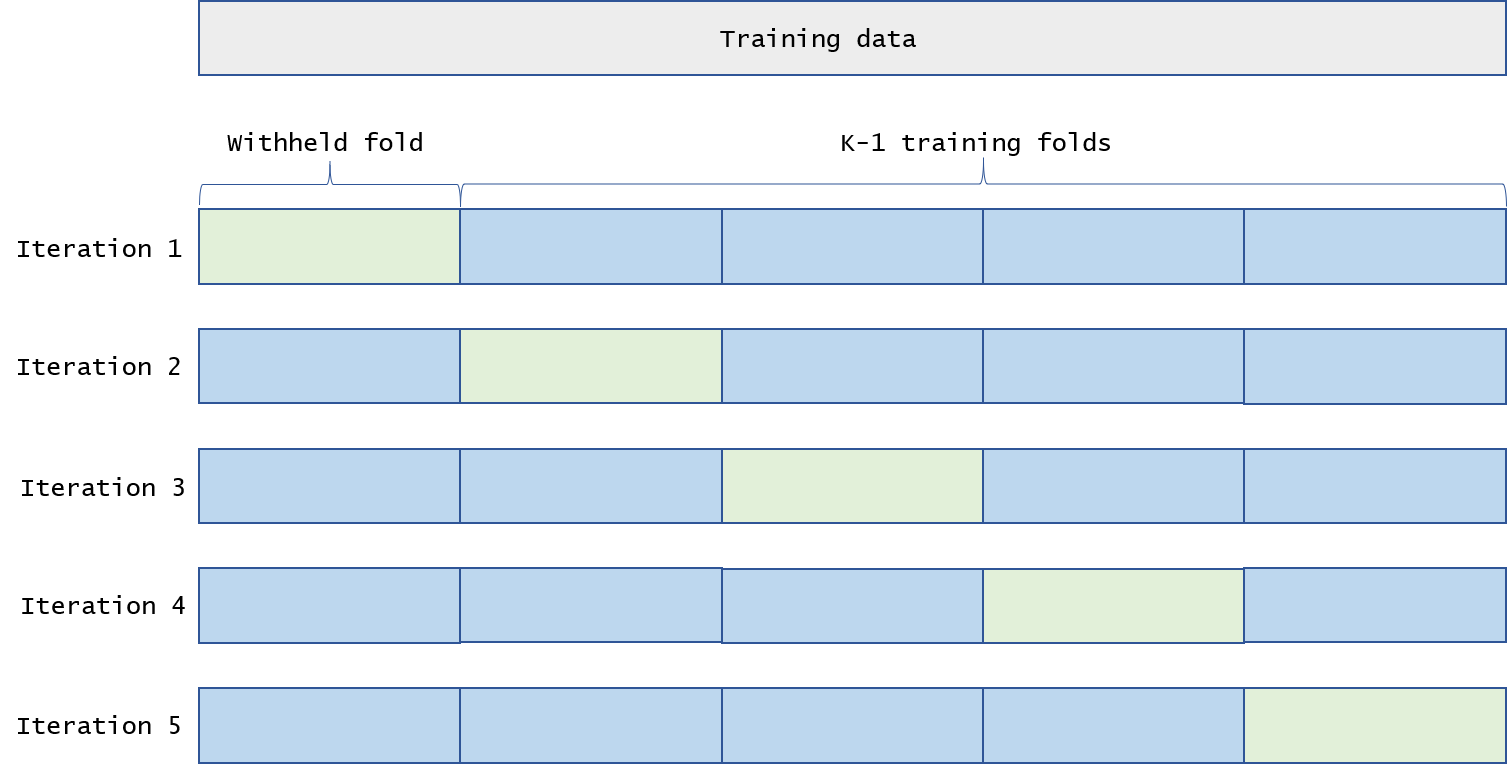}
\caption{$k$-fold cross-validation. We represent the training data as a light grey rectangle, and we illustrate $5$-fold cross-validation on the training data. In each iteration, the withheld fold is colored light green and the $4$ training folds are colored light blue. We note that each of the withheld folds is used exactly once as the validation data.}
\label{fig:cross_validation}
\end{figure}

In Fig.~\ref{fig:cross_validation}, we illustrate $k$-fold cross-validation, e.g., $k=5,$ which is a precursor for nested cross-validation~\cite{Hastie_2009, Cawley_2010, Sklearn_2011} and the inbuilt model selection step in Alg.~\ref{alg:2}. The method begins by randomly partitioning Eq.~\ref{eq:transformed_training_data} into $k$ non-overlapping folds, and $k$ is typically a natural number between $5$ and $10,$ inclusive.
Next, we repeat the following two steps $k$ times with each of the withheld folds used exactly once as the validation data:\
\begin{itemize}
\item Of the $k$ folds, we withhold one for validation. A single-target boosting algorithm receives the remaining $k-1$ folds as training data, and we request a single-target regressor as output.
\item We evaluate the induced single-target regressor on the withheld fold from the previous step by computing the average loss of the single-target regressor.
\end{itemize}
Then, we average the $k$ results from the second step, and we refer to this average as cross-validation error. This completes a single loop of the $k$-fold cross-validation method. In best practices of machine learning, this method is preferred over leave-one-out cross-validation, wherein $k=m_{train}$~\cite{Hastie_2009, Cawley_2010, Sklearn_2011}.

In nested cross-validation, the estimation method utilizes an outer loop of $k$ non-overlapping folds and an inner loop of $l$ non-overlapping folds. The outer loop is utilized to estimate the $j$th mean absolute error Eq.~\ref{eq:single_target_test_error} and the inner loop is utilized to select the (hyper)parameters in Alg.\ \ref{alg:1}, such as the choice of basis function or value of $K_{j}$ in Eq.~\ref{eq:additive_model}. The method begins by randomly partitioning Eq.~\ref{eq:transformed_training_data} into $k$ non-overlapping folds. Next, we repeat the following two steps $k$ times with each of the withheld folds in the outer loop used exactly once as the validation data:\
\begin{itemize}
\item Of the $k$ folds, we withhold a fold for validation. In the inner loop, we apply $l$-fold cross-validation to the remaining $k-1$ folds for multiple single-target boosting algorithms with differing (hyper)parameters. After completing the inner loop, we select the best performing single-target boosting algorithm based on the minimum inner loop cross-validation error.
\item The selected single-target boosting algorithm receives the $k-1$ folds from the previous step as training data, and we request a single-target regressor as output. We evaluate the induced single-target regressor on the withheld fold from the previous step by computing the average loss of the single-target regressor.
\end{itemize}
Then, we average the $k$ results from the second step, and we use this average to approximate the $j$th mean absolute error Eq.~\ref{eq:single_target_test_error}. In practice, we usually execute nested cross-validation within an exhaustive hyperparameter search tool, such as $GridSearchCV$ by scikit-learn~\cite{Sklearn_2011} (see Supplementary Information for implementation details).  

After completing nested cross-validation, we repeat the second step in the learning framework. In the $j$th single-target regression subtask, Alg.~\ref{alg:2} utilizes Eq.~\ref{eq:transformed_training_data} in a modified $k$-fold cross-validation procedure to select either the incumbent smooth term from the base regressor or the candidate additive expansion Eq.~\ref{eq:additive_model} as the induced single-target regressor. This entails modifying the second step in the aforedescribed $k$-fold cross-validation method, namely
\begin{itemize}
\item We independently evaluate the smooth term and the induced single-target regressor on the withheld fold from the previous step by computing the average loss of the smooth term and the single-target regressor. We note that the smooth term always predicts the $j$th feature, given an example from the withheld fold.
\end{itemize}
Then, we independently average their $k$ results, and we refer to these averages as the incumbent error and the cross-validation error, respectively. The inbuilt model selection step in Alg.~\ref{alg:2} selects the better algorithm based on the minimum error. Subsequently, the boosting algorithm completes each single-target regression subtask, and the boosting algorithm returns the induced multi-target regressor for evaluation on the test data.

\textbf{Single-target gradient boosting} -- For the $j$th single-target regression subtask, the single-target boosting algorithm Alg.~\ref{alg:1} takes as input training examples Eq.~\ref{eq:transformed_training_data}, number of iterations $K_{j},$ single-target loss functions $\{\ell_{j}, \tilde{\ell}_{j}\},$ and basis function $b$ characterized by parameter set $\theta.$ 
\begin{figure*}
\begin{minipage}{\linewidth}
\begin{algorithm}[H]
\SetAlgoLined
\SetKwInOut{Input}{Input}
\Input{training examples\\
number of iterations $K_{j}$ \\
single-target loss functions $\{\ell_{j}, \tilde{\ell}_{j}\}$ \\
basis function $b$ characterized by parameter set $\theta$}
\SetKwInput{Initialize}{Initialize}
\Initialize{$h_{j, 0}(X) = X_{j}$ for each example}
\For{$k=1$ \KwTo $K_{j}$}{
\begin{enumerate}
\item[(a)]
\For{$i=1$ \KwTo $m_{train}$}{
Compute pseudo-residuals \\
\qquad $r^{(i)}_{j,k} = - \frac{\partial \ell_{j}(Y_{j}^{(i)}, h_{j}(X^{(i)}))}{\partial h_{j}(X^{(i)})} \bigg|_{h_{j}(X^{(i)})=h_{j, k-1}(X^{(i)})}$}
\item[(b)] Induce a basis function on $\{X^{(i)}, r_{j,k}^{(i)}\}_{i=1}^{m_{train}}$ to learn the parameter set $\theta_{j,k}$
\item[(c)] Solve the one-dimensional optimization problem to learn the expansion coefficient\\
$\alpha_{j, k} = \argmin_{\alpha} \sum_{i=1}^{m_{train}} \tilde{\ell}_{j} (Y^{(i)}_{j}, h_{j, k-1}(X^{(i)}) +\alpha b(X^{(i)}; \theta_{j, k}))$
\item[(d)] Sequentially append the induced basis function to the additive expansion \\
$h_{j, k}(X) = h_{j, k-1}(X) + \alpha_{j, k} b(X; \theta_{j, k})$
\end{enumerate}
}
\SetKwInOut{Output}{Output}
\Output{Single-target regressor $\hat{h}_{j} = h_{j, K_{j}}$}
\caption{$BaseBoost$}
\label{alg:1}
\end{algorithm}
\end{minipage}
\end{figure*}
For instance, the parameter set would encode the split features, split locations, and the terminal node means of the individual trees, if the choice of basis function were a shallow decision tree; see for example~\cite{Friedman_2001, Friedman_2003, Hastie_2009, Chen_2016, Ke_2017}. In the application, we choose a stacking regressor~\cite{Wolpert_1992, Breiman_1996, Sklearn_2011} as the basis function, which is a two layer ensemble of single-target regressors (see Supplementary Information). 

In Alg.~\ref{alg:1}, the first line initializes to the smooth term for each example in Eq.~\ref{eq:transformed_training_data}. In the for loop, line~(a) computes the pseudo-residuals with single-target loss function $\ell_{j},$ whereby the term pseudo-residual emanates from the term residual in least squares fitting and reroughing~\cite{Tukey_1977, Friedman_2001, Friedman_2003, Hastie_2009}. Line~(b) enables the boosting algorithm to work for any given single-target learning algorithm~\cite{Friedman_2001, Friedman_2003, Hastie_2009}, whereby the labels are the pseudo-residuals from line~(a). Line~(c) computes the one-dimensional line search with single-target loss function $\tilde{\ell}_{j}.$ Line~(d) sequentially appends the basis function to the additive expansion. The output is the induced single-target regressor $\hat{h}_{j}$.

In the application, we modify line~(c) in Alg.~\ref{alg:1} to include $\mbox{L}^{1}$ regularization (see the Supplementary Information). In relation to previous work, the initialization step in Alg.~\ref{alg:1} depends upon the examples, whereas the standard form of gradient boosting initializes to the optimal constant model:\
$\argmin_{c} \sum_{i=1}^{m_{train}} \ell_{j}(Y_{j}^{(i)}, c);$ see references~\cite{Friedman_2001, Friedman_2003, Hastie_2009}. In matching pursuit and its extensions, the greedy stagewise algorithms initialize to the zero vector, and they sequentially transform the signal into a negligible residual; see references~\cite{Mallat_1993, Vincent_2002, Donoho_2012} for the algorithmic body differences and further details. 

For the $j$th single-target regression subtask, the augmented version of the single-target boosting algorithm Alg.~\ref{alg:2} takes as input training examples Eq.~\ref{eq:transformed_training_data}, number of iterations $K_{j},$ single-target loss functions $\{\ell_{j}, \tilde{\ell}_{j}\},$ basis function $b$ characterized by parameter set $\theta,$ number of cross-validation folds $k,$ and $k$-fold cross-validation single-target loss function. 
\begin{figure*}
\begin{minipage}{\linewidth}
\begin{algorithm}[H]
\SetAlgoLined
\SetKwInOut{Input}{Input}
\Input{training examples\\
number of iterations $K_{j}$ \\
single-target loss functions $\{\ell_{j}, \tilde{\ell}_{j}\}$ \\
basis function $b$ characterized by parameter set $\theta$ \\
number of cross-validation folds $k$\\
$k$-fold cross-validation single-target loss function}
\SetKwInput{Selection}{Model selection}
\Selection{Perform modified $k$-fold cross-validation for the incumbent smooth term and the candidate $BaseBoost.$ If the incumbent error is less than or equal to the cross-validation error, then break $\hat{h}_{j}(X) = X_{j}.$ Otherwise, call $BaseBoost.$}
\SetKwInOut{Output}{Output}
\Output{Single-target regressor $\hat{h}_{j}$}
\caption{$BaseBoostCV$}
\label{alg:2}
\end{algorithm}
\end{minipage}
\end{figure*}
The inbuilt model selection step in Alg.~\ref{alg:2} selects the incumbent smooth term as the induced single-target regressor, if the incumbent error is less than or equal to the cross-validation error, otherwise Alg.~\ref{alg:2} calls Alg.~\ref{alg:1} ($k$-fold cross-validation details in previous section). The output is the induced single-target regressor $\hat{h}_{j}$. 

\textbf{Explainable machine learning} -- In machine learning competitions and products, complex models, such as ensemble and deep learning models, are omnipresent. Understanding why these models make certain predictions is the focus of explainable machine learning~\cite{Lundberg_2017, Molnar_2020}. The SHAP framework unifies several approaches in explainable machine learning to replicate individual predictions generated by a single-target regressor with a simpler linear explanation model whose coefficients measure feature importance~\cite{Lundberg_2017}. In the work of \u{S}trumbelj and Kononenko, these coefficients, known as SHAP values~\cite{Lundberg_2017}, were shown to be equivalent to the Shapley value in cooperative game theory~\cite{Strumbelj_2014}. The explanation model is defined as a linear function of binary variables
\begin{equation}
\label{eq:explanation_model}
g(z') = \phi_{0} + \sum_{k=1}^{M} \phi_{k} z'_{k}
\end{equation}
where $z' \in \{ 0, 1\}^{M}$ is a set of binary variables, $M$ is the number of features under consideration, and $\phi_{k}$ is a real-valued feature attribution, known as a SHAP value, for the $k$th feature. As the computation of Shapley values has an exponential time complexity~\cite{Strumbelj_2014}, the SHAP software approximates the coefficients with insights from additive feature attribution methods; see~\cite{Lundberg_2017}. 

In the application, we utilize the model-agnostic approximation method, known as Kernel SHAP, to compute the SHAP values~\cite{Lundberg_2017}. This enables us to ascertain a simpler explanation model to approximate each training data prediction generated by the induced fully-connected neural networks~\cite{Sklearn_2011}, where $M=9$ in Eq.~\ref{eq:explanation_model} for the control voltage features (see the Supplementary Information). The importance $I$ of each feature is defined as the sum of absolute SHAP values 
\begin{equation}
\label{eq:feature_importance}
I_{k} = \sum_{i=1}^{m_{train}} |\phi_{k}^{(i)}|,
\end{equation}
which enables an ordering to be defined. The features are sorted in ascending order from bottom to top in each summary plot~\cite{Lundberg_2017, Molnar_2020}.

\section*{Data Availability}

All data, relevant to the information and figures presented in this manuscript, are available upon reasonable request. 

\section*{Author Contributions}

A.W.\ designed the learning approach, implemented the learning system, and performed the data analysis. All authors contributed to the interpretation of the data and to writing the manuscript.

\bibliographystyle{unsrt}
\bibliography{references}

\begin{thebibliography}{10}

\bibitem{Hume_1739}
David Hume.
\newblock {\em A Treatise of Human Nature}.
\newblock Clarendon Press, 1739.

\bibitem{Haussler_1988}
David Haussler.
\newblock Quantifying inductive bias: {AI} learning algorithms and {V}aliant's
  learning framework.
\newblock {\em Artificial Intelligence}, 36:177--221, 1988.

\bibitem{Mitchell_1991}
Tom Mitchell.
\newblock The need for biases in learning generalisation.
\newblock In Jude Shavlik and Thomas Dietterich, editors, {\em Readings in
  Machine Learning}. Morgan Kaufmann, 1991.

\bibitem{Wolpert_1997}
David Wolpert and William Macready.
\newblock No free lunch theorems for optimization.
\newblock {\em IEEE Transactions on Evolutionary Computation}, 1(1), 1997.

\bibitem{Caruana_1997}
Rich Caruana.
\newblock Multitask learning.
\newblock {\em Machine Learning}, 28:41--75, 1997.

\bibitem{Baxter_2000}
Jonathan Baxter.
\newblock A model of inductive bias learning.
\newblock {\em Journal of Artificial Intelligence Research}, 12:149--198, 2000.

\bibitem{Brunton_2016}
Steven Brunton, Joshua Proctor, and Jos\'{e}~Nathan Kutz.
\newblock Discovering governing equations from data by sparse identification of
  nonlinear dynamical systems.
\newblock {\em Proceedings of the National Academy of Sciences of the United
  States of America}, 113(15):3932--3937, 2016.

\bibitem{Roushan_2017}
Pedram~Roushan et~al.
\newblock Spectroscopic signatures of localization with interacting photons in
  superconducting qubits.
\newblock {\em Science}, 358:1175--1179, 2017.

\bibitem{Butler_2018}
Keith~Butler et~al.
\newblock Machine learning for molecular and materials science.
\newblock {\em Nature}, 559:547--555, 2018.

\bibitem{Neill_2018}
Charles~Neill et~al.
\newblock A blueprint for demonstrating quantum supremacy with superconducting
  qubits.
\newblock {\em Science}, 360:195--199, 2018.

\bibitem{Chiaro_2019}
Benjamin~Chiaro et~al.
\newblock Growth and preservation of entanglement in a many-body localized
  system.
\newblock arXiv:1910.06024, 2019.

\bibitem{Schmidt_2009}
Michael Schmidt and Hod Lipson.
\newblock Distilling free-form natural laws from experimental data.
\newblock {\em Science}, 324:81--85, 2009.

\bibitem{Carrasquilla_2017}
Juan Carrasquilla and Roger Melko.
\newblock Machine learning phases of matter.
\newblock {\em Nature Physics}, 13:431--434, 2017.

\bibitem{Melnikov_2018}
Alexey~Melnikov et~al.
\newblock Active learning machine learns to create new quantum experiments.
\newblock {\em Proceedings of the National Academy of Sciences of the United
  States of America}, 115(6):1221--1226, 2018.

\bibitem{Koch-Janusz_2018}
Maciej Koch-Janusz and Zohar Ringel.
\newblock Mutual information, neural networks and the renormalization group.
\newblock {\em Nature Physics}, 14:578--582, 2018.

\bibitem{Torlai_2018}
Giacomo~Torlai et~al.
\newblock Neural-network quantum state tomography.
\newblock {\em Nature Physics}, 14:447--450, 2018.

\bibitem{Wu_2019}
Tailin Wu and Max Tegmark.
\newblock Toward an artificial intelligence physicist for unsupervised
  learning.
\newblock {\em Physical Review E}, 100:033311, 2019.

\bibitem{Iten_2020}
Raban~Iten et~al.
\newblock Discovering physical concepts with neural networks.
\newblock {\em Physical Review Letters}, 124:010508, 2020.

\bibitem{Wetzel_2020}
Sebastian~Wetzel et~al.
\newblock Discovering symmetry invariants and conserved quantities by
  interpreting siamese neural networks.
\newblock arXiv:2003.04299, 2020.

\bibitem{Borchani_2015}
Hanen~Borchani et~al.
\newblock A survey on multi-output regression.
\newblock {\em Wiley Interdisciplinary Reviews:\ Data Mining and Knowledge
  Discovery}, 2015.

\bibitem{Waegeman_2019}
Willem Waegeman, Krzysztof Dembczy\'{n}ski, and Eyke H\"{u}llermeier.
\newblock Multi-target prediction: A unifying view on problems and methods.
\newblock {\em Data Mining and Knowledge Discovery}, 33:293--324, 2019.

\bibitem{Sklearn_2011}
Fabian~Pedregosa et~al.
\newblock Scikit-learn:\ machine learning in python.
\newblock {\em Journal of Machine Learning Research}, 12:2825--2830, 2011.

\bibitem{Pascual_leone_2005}
Alvaro Pascual-Leone et~al.
\newblock The plastic human brain cortex.
\newblock {\em Annual Review of Neuroscience}, 28:377--401, 2005.

\bibitem{Chen_2014}
Yu~Chen et~al.
\newblock Qubit architecture with high coherence and fast tunable coupling.
\newblock {\em Physical Review Letters}, 113:220502, 2014.

\bibitem{Zahedinejad_2016}
Ehsan Zahedinejad, Joydip Ghosh, and Barry Sanders.
\newblock Designing high-fidelity single-shot three-qubit gates:\ a
  machine-learning approach.
\newblock {\em Physical Review Applied}, 6:054005, 2016.

\bibitem{Lin_2017}
Henry Lin, Max Tegmark, and David Rolnick.
\newblock Why does deep and cheap learning work so well?
\newblock {\em Journal of Statistical Physics}, 168(6):1223--1247, 2017.

\bibitem{Biamonte_2017}
Jacob~Biamonte et~al.
\newblock Quantum machine learning.
\newblock {\em Nature}, 549:195--202, 2017.

\bibitem{Dunjko_2018}
Vedran Dunjko and Hans Briegel.
\newblock {Machine Learning {\&} Artificial Intelligence in the Quantum Domain:
  A Review of Recent Progress}.
\newblock {\em Reports on Progress in Physics}, 81(7), 2018.

\bibitem{Giuseppe_2019}
Giuseppe~Carleo et~al.
\newblock Machine learning and the physical sciences.
\newblock {\em Reviews of Modern Physics}, 91:045002, 2019.

\bibitem{Mehta_2019}
Pankaj~Mehta et~al.
\newblock {A high-bias, low-variance introduction to Machine Learning for
  physicists}.
\newblock {\em Physics Reports}, 810:1--124, 2019.

\bibitem{Udrescu_2020}
Silviu-Marian Udrescu and Max Tegmark.
\newblock {AI} {Feynman}:\ {A} physics-inspired method for symbolic regression.
\newblock {\em Science Advances}, 6(16), 2020.

\bibitem{Lundberg_2017}
Scott Lundberg and Su-In Lee.
\newblock A unified approach to interpreting model predictions.
\newblock In {\em Advances in Neural Information Processing Systems 30}, pages
  4765--4774. Curran Associates, Inc., 2017.

\bibitem{Molnar_2020}
Christoph Molnar.
\newblock Interpretable machine learning:\ a guide for making black box models
  explainable.
\newblock christophm.github.io/interpretable-ml-book, 2020.

\bibitem{Linke_2017}
Norbert~Linke et~al.
\newblock Experimental comparison of two quantum computing architectures.
\newblock {\em Proceedings of the National Academy of Sciences of the United
  States of America}, 114(13):3305--3310, 2017.

\bibitem{Kelly_2018_1}
Julian~Kelly et~al.
\newblock Physical qubit calibration on a directed acyclic graph.
\newblock arXiv:1803.03226, 2018.

\bibitem{Wolpert_1992}
David Wolpert.
\newblock Stacked generalization.
\newblock {\em Neural Networks}, 5:241--259, 1992.

\bibitem{Breiman_1996}
Leo Breiman.
\newblock Stacked regressions.
\newblock {\em Machine Learning}, 24:49--64, 1996.

\bibitem{Breiman_1997_2}
Leo Breiman and Jerome Friedman.
\newblock Predicting multivariate responses in multiple linear regression.
\newblock {\em Royal Statistical Society Series B}, 59:3--54, 1997.

\bibitem{Erhan_2010}
Dumitru~Erhan et~al.
\newblock Why does unsupervised pre-training help deep learning?
\newblock {\em Journal of Machine Learning Research}, 11:625--660, 2010.

\bibitem{Haussler_1992}
David Haussler.
\newblock Decision theoretic generalizations of the {PAC} model for neural net
  and other learning applications.
\newblock {\em Information and Computation}, 100:78--150, 1992.

\bibitem{Kearns_1994_1}
Michael Kearns and Robert Schapire.
\newblock Efficient distribution-free learning of probabilistic concepts.
\newblock {\em Journal of Computer and Systems Science}, 48:464--497, 1994.

\bibitem{Kearns_1994_2}
Michael Kearns, Robert Schapire, and Linda Sellie.
\newblock Toward efficient agnostic learning.
\newblock {\em Machine Learning}, 17:115--141, 1994.

\bibitem{Friedman_2001}
Jerome Friedman.
\newblock Greedy function approximation: A gradient boosting machine.
\newblock {\em Annals of Statistics}, 29(5):1189--1232, 2001.

\bibitem{Friedman_2003}
Jerome Friedman and Bogdan Popescu.
\newblock Importance sampled learning ensembles.
\newblock Technical report, Stanford University, Department of Statistics,
  2003.

\bibitem{Hastie_2009}
Trevor Hastie, Robert Tibshirani, and Jerome Friedman.
\newblock {\em The Elements of Statistical Learning}.
\newblock Springer, 2009.

\bibitem{Ng_2020}
Andrew Ng.
\newblock Machine learning yearning.
\newblock deeplearning.ai project, 2020.

\bibitem{Powell_1987}
Michael Powell.
\newblock Radial basis functions for multivariable interpolation:\ a review.
\newblock In {\em Algorithms for Approximation}. Clarendon Press, 1987.

\bibitem{Kearns_1988}
Michael Kearns and Leslie Valiant.
\newblock Learning boolean formulae or finite automata is as hard as factoring.
\newblock Technical Report TR-14-88, Harvard University Aiken Computation
  Laboratory, 1988.

\bibitem{Hastie_1990}
Trevor Hastie and Robert Tibshirani.
\newblock {\em Generalized Additive Models}.
\newblock Chapman and Hall, London, 1990.

\bibitem{Schapire_1990}
Robert Schapire.
\newblock The strength of weak learnability.
\newblock {\em Machine Learning}, 5:197--227, 1990.

\bibitem{Kearns_1994_3}
Michael Kearns and Leslie Valiant.
\newblock Cryptographic limitations on learning boolean formulae and finite
  automata.
\newblock {\em Journal of the Association for Computing Machinery}, 41:67--95,
  1994.

\bibitem{Freund_1995}
Yoav Freund.
\newblock Boosting a weak learning algorithm by majority.
\newblock {\em Information and Computation}, 121:256--285, 1995.

\bibitem{Freund_1997}
Yoav Freund and Robert Schapire.
\newblock A decision-theoretic generalization of on-line learning and an
  application to boosting.
\newblock {\em Journal of Computer and System Sciences}, 55:119--139, 1997.

\bibitem{Breiman_1997_1}
Leo Breiman.
\newblock Arcing the edge.
\newblock Technical report, Stanford University, Department of Statistics,
  1997.

\bibitem{Mason_1999}
Llew~Mason et~al.
\newblock Boosting algorithms as gradient descent.
\newblock In {\em NIPS:\ Proceedings of the $12$th International Conference on
  Neural Information Processing}, pages 512--–518, 1999.

\bibitem{Schapire_2002}
Robert~Schapire et~al.
\newblock Incorporating prior knowledge into boosting.
\newblock In {\em ICML'02:\ Proceedings of the Nineteenth International
  Conference on Machine Learning}, pages 538--545, 2002.

\bibitem{Chen_2016}
Tianqi Chen and Carlos Guestrin.
\newblock {XGBoost}:\ a scalable tree boosting system.
\newblock In {\em KDD'16:\ Proceedings of the $22$nd ACM SIGKDD International
  Conference on Knowledge Discovery and Data Mining}, pages 785--794, 2016.

\bibitem{Ke_2017}
Guolin ke~et al.
\newblock {LightGBM}:\ a highly efficient gradient boosting decision tree.
\newblock In {\em Advances in Neural Information Processing Systems 30}, pages
  3149--3157. Curran Associates, Inc., 2017.

\bibitem{Tukey_1977}
John Tukey.
\newblock {\em Exploratory Data Analysis}.
\newblock Addison-Wesley, 1977.

\bibitem{Friedman_2000}
Jerome Friedman, Trevor Hastie, and Robert Tibshirani.
\newblock Additive logistic regression: A statistical view of boosting.
\newblock {\em The Annals of Statistics}, 28(2):337--407, 2000.

\bibitem{Mallat_1993}
St\'{e}phane Mallat and Zhifeng Zhang.
\newblock Matching pursuits with time-frequency dictionaries.
\newblock {\em IEEE Transactions on Signal Processing}, 41(12):3397--3415,
  1993.

\bibitem{Vincent_2002}
Pascal Vincent and Yoshua Bengio.
\newblock Kernel matching pursuit.
\newblock {\em Machine Learning}, 48:165--187, 2002.

\bibitem{Donoho_2012}
David~Donoho et~al.
\newblock Sparse solution of underdetermined systems of linear equations by
  stagewise orthogonal matching pursuit.
\newblock {\em IEEE Transactions on Information Theory}, 58(2):1094--1121,
  2012.

\bibitem{He_2014}
Xinran~He et~al.
\newblock Practical lessons from predicting clicks on ads at facebook.
\newblock In {\em ADKDD'14:\ Proceedings of the Eighth International Workshop
  on Data Mining for Online Advertising}, 2014.

\bibitem{Shalizi_2001}
Cosma Shalizi and James Crutchfield.
\newblock Computational mechanics:\ {Pattern} and prediction, structure and
  simplicity.
\newblock {\em Journal of Statistical Physics}, 104(3-4):817--879, 2001.

\bibitem{Gu_2012}
Mile~Gu et~al.
\newblock Quantum mechanics can reduce the complexity of classical models.
\newblock {\em Nature communications}, 3(1):1--5, 2012.

\bibitem{Cawley_2010}
Gavin Cawley and Nicola Talbot.
\newblock On over-fitting in model selection and subsequent selection bias in
  performance evaluation.
\newblock {\em Journal of Machine Learning Research}, 11:2079--2107, 2010.

\bibitem{Strumbelj_2014}
Erik \u{S}trumbelj and Igor Kononenko.
\newblock Explaining prediction models and individual predictions with feature
  contributions.
\newblock {\em Knowledge and Information Systems}, 41:647--665, 2014.

\end{thebibliography}
\clearpage
\onecolumngrid


\section*{Supplementary material}

# Supplementary Information: Boosting on the shoulders of giants in quantum device calibration

Alex Wozniakowski,[1,2,*] Jayne Thompson,[3] Mile Gu,[1,2,†] and Felix Binder[4]
[1]*School of Physical and Mathematical Sciences, Nanyang Technological University*
[2]*Complexity Institute, Nanyang Technological University*
[3]*Centre for Quantum Technologies, National University of Singapore*
[4]*Institute for Quantum Optics and Quantum Information – IQOQI Vienna,
Austrian Academy of Sciences, Boltzmanngasse 3, 1090 Vienna, Austria*

## S1. BENCHMARK DATASET

In this section, we briefly review the quantum device and the classical control program utilized in the benchmark task, and we describe the contents in the benchmark dataset. For a detailed introduction to the many-body Ramsey spectroscopy technique, the classical control program, calibration methodologies, and the superconducting quantum device, see for example [S1–S5].

The quantum device is a nearest-neighbor coupled linear chain of 9 superconducting qubits, wherein the 5 rightmost qubits and 4 interceding couplings were utilized during the benchmark dataset acquisition [S1–S3, S5]; see the main body of reference [S2] for an optical micrograph of the device and see the Supplementary Materials of reference [S3] for the corresponding electronic circuit diagram. Each qubit is explicitly modeled as a capacitor, inductor, and tunable junction, all in series (see the Supplementary Materials of reference [S3] for the physical parameters utilized in the classical control program. For the device architecture, there are 26 control lines used to drive the microwave rotations, set the qubit frequencies, and bias the couplers [S3].

To obtain each multi-target in the benchmark dataset, an instance of the many-body Ramsey spectroscopy technique [S2] begins by setting the parameters in Eq. 1 such that the on-site detuning is sampled uniformly in $[-100, 100]$ MHz, the hopping rate is sampled uniformly in $[0, 50]$ MHz, and the on-site Hubbard interaction is fixed at 0. Next, the time-domain spectroscopy circuit in Fig. S1 is run 5 times with these parameters, and we denote the choice of superposition qubit and readout resonator by the index $k$, where $k \in \{1, 2, \ldots, 5\}$.

In the $k$th run of the time-domain spectroscopy circuit, each qubit starts in the fiducial state $|0\rangle$, and no photon is present in the system. Next, a microwave pulse is applied to the $k$th qubit, e.g., $k = 1$ in Fig. S1, which places the qubit in a superposition of the computational basis and it initializes a single-photon in the system. Then, the system evolves according to the time-independent Hamiltonian Eq. 1. After the evolution, a microwave pulse is applied to the $k$th qubit to measure

---
* wozn0001@e.ntu.edu.sg
† mgu@quantumcomplexity.org

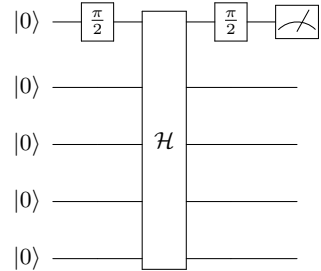

FIG. S1. Time-domain spectroscopy circuit. Initially, each qubit is in the fiducial state $|0\rangle$. Using a microwave pulse, the $k$th specified qubit, e.g., $k = 1$, is placed in a superposition of the computational basis. Next, the system evolves according to the time-independent Hamiltonian Eq. 1 with randomly programmed parameters. After the evolution, a microwave pulse is applied to measure either $\langle \sigma^X \rangle$ or $\langle \sigma^Y \rangle$, where $\sigma^X$ and $\sigma^Y$ denote Pauli operators.

either $\langle \sigma^X \rangle$ or $\langle \sigma^Y \rangle$. From the measurement of these observables, the observable $\langle \sigma^X \rangle + i \langle \sigma^Y \rangle$ is instantiated; and the energy spectrum is fully resolved by completing all 5 runs [S2]. We note that the choice of the operator is designed to isolate the single-photon manifold, so all of the eigenenergies have the same sign (see the Supplementary Materials in reference [S2]). Lastly, the peaks in the fast Fourier transform of the observable $\langle \sigma^X \rangle + i \langle \sigma^Y \rangle$ are identified as the eigenenergies of the Hamiltonian [S2, S5], and these eigenenergies are sorted into ascending order, as described in the main body.

To obtain each multi-target prediction in the benchmark dataset, the classical control program [S2, S3, S5] maps a collection of 5 qubit and 4 coupler bias features to the $5 \times 5$ single-photon block matrix in the representation of Eq. 1. Next, a numerical eigensolver produces 5 eigenenergy approximations, and they are sorted in ascending order.

In the benchmark dataset, there are $m = 136$ of each: qubit and coupler bias examples, multi-target predictions generated by the control program, and multi-targets retrieved by the many-body Ramsey spectroscopy technique [S2]. In splitting this data for machine learning, we want to preserve the experimental association, i.e., the examples form a triple consisting of 5 qubit bias features, 4 coupler bias features, 5 single-target predictions, and 5 single-targets. Moreover, we split these examples into $m_{train} = 95$ and $m_{test} = 41$ indices for training and

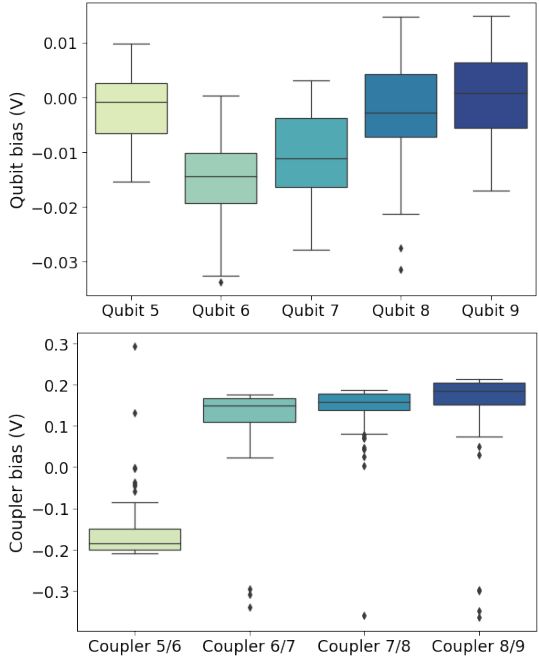

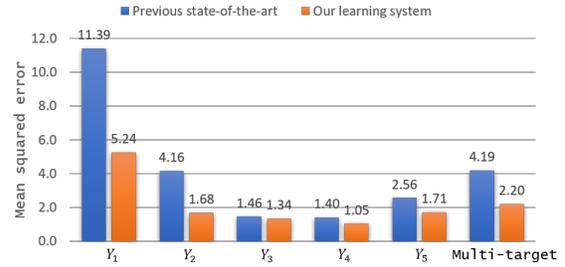

FIG. S4. Using the squared error loss function as the criterion, our learning system surpasses the previous state-of-the-art [S2, S3, S5] by over 47% on the multi-target regression task. Moreover, our learning system outperforms the previous state-of-the-art [S2, S3, S5] on each single-target regression subtask.

## S2. TEST ERROR

In the application, the $j$th single-target test error is mean absolute error Eq. 8, and the multi-target test error is average mean absolute error Eq. 9, where $n = 5$ and $m_{test} = 41$. In the main body, we plot the mean absolute errors and average mean absolute error of the previous state-of-the-art [S2, S3, S5] and our learning system in Fig. 2.

In the real world application of machine learning, multiple measures of prediction performance should be studied and reported [S6–S8]. Here, we utilize the squared error loss function to measure the prediction performance in the application, as we have already generated the test example predictions. Importantly, the squared error loss function is decomposable over the targets, namely

$$\ell(Y, h(X)) = ||Y - h(X)||_2^2 \\ = \sum_{j=1}^{n}(Y_j - h_j(X))^2 \quad \text{(S1)} \\ = \sum_{j=1}^{n}\ell_j(Y_j, h_j(X)).$$

The $j$th single-target test error is mean squared error

$$\frac{1}{m_{test}}\sum_{i=1}^{m_{test}}(Y_j^{(i)} - h_j(X^{(i)}))^2, \quad \text{(S2)}$$

and the multi-target test error is the average mean squared error

$$\frac{1}{n}\sum_{j=1}^{n}\frac{1}{m_{test}}\sum_{i=1}^{m_{test}}(Y_j^{(i)} - h_j(X^{(i)}))^2. \quad \text{(S3)}$$

From left to right in Fig. S4, we show the mean squared errors Eq. S2 and the average mean squared error Eq. S2 of the previous state-of-the-art [S2, S3, S5] and our learning system, in blue and orange, respectively. Our

FIG. S2. Qubit and coupler bias feature boxplots. We depict the qubit bias features utilized in the training data split (top), whereby the label Qubit $j$ denotes the qubit bias corresponding to qubit site $j$. We depict the coupler bias features utilized in the training data split (bottom), whereby the label Coupler $j/j+1$ denotes the coupler bias corresponding to the nearest neighbor coupler for qubit sites $j$ and $j + 1$.

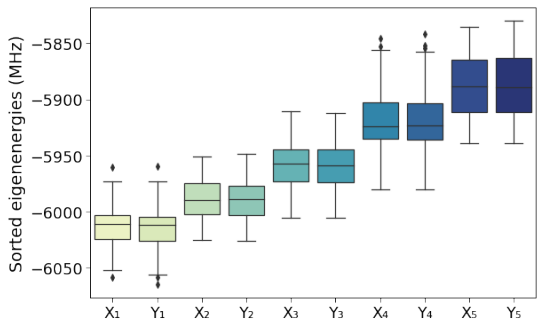

FIG. S3. Joint single-target prediction feature and single-target observation boxplot. We jointly depict the single-target predictions and the single-target observations utilized in the training data split.

test data, respectively. In the boxplots in Fig. S2, we depict the qubit bias features utilized in the training data split (top), and we depict the coupler bias features utilized in the training data split (bottom). In the boxplot in Fig. S3, we jointly depict the single-target predictions and the single-target observations utilized in the training data split.

learning system outperforms the previous state-of-the-art [S2, S3, S5] on each single-target regression subtask and the multi-target regression task. Moreover, our learning system surpasses the previous state-of-the-art [S2, S3, S5] by over 47% on the multi-target regression task.

## S3. MULTI-TARGET STACKING

Multi-target regression applications pose novel research questions and there is a demand for new methods, which consider not only the underlying relationships between the features and the associated single-targets but also the relationships between the single-targets [S9, S10]. Here, we review the single-target approach, which does not consider the relationship between the single-targets, as well as the prototypical two-step stacking approach, which addends a second layer to the single-target approach in order to exploit dependencies among the single-targets.

In the single-target approach, a learning algorithm receives training examples and we request a multi-target regressor as output. For each single-target regression subtask, the learning algorithm independently induces a single-target regressor on an appropriate slice of the training data [S9, S10]. After finishing the subtasks, the learning algorithm concatenates each single-target regressor into a multi-target regressor. Given an example, the multi-target regressor predicts a real vector. As many learning algorithms do not natively support multi-target prediction, this approach is widely applied in multi-target regression applications [S8–S10]. However, this approach does not imply simpler single-target regressors than an approach which considers the relationship between the single-targets [S9, S10]. This leads to the algorithm design question: can we do better by exploiting dependencies among the single-targets?

Initially introduced in single-target classification [S11] and regression tasks [S12], stacking is a general ensemble learning technique for combining single-target classifiers or regressors to reduce their biases. In the multi-target regression setting, the two-step stacking approach enforces the idea that single-target regressors should behave similarly in order to outperform the independent single-target approach [S9, S10, S13].

In the two-step stacking approach, a learning algorithm receives training examples and we request a multi-target regressor as output. In the first step, the learning algorithm applies the single-target approach, and it uses the induced multi-target regressor to generate training example predictions. In the second step, the learning algorithm regards these predictions as new examples while retaining the initial labels. In some variations, the learning algorithm regards a concatenation of the original examples with these predictions as new examples while retaining the initial labels [S9, S10]. Next, the learning algorithm applies the single-target approach to induce a second layer of single-target regressors, and the learning algorithm returns a composition of the first and second layer multi-target regressors. In other words, given an example from the domain, the first layer multi-target regressor generates an initial prediction, then the second layer multi-target regressor regularizes this prediction, which reduces the problem of overfitting [S9, S10, S13]. This technique is related to methods in deep learning, such as pre-training [S14] and weight sharing [S10, S15]. Our learning framework is motivated by this two-step stacking approach [S9–S13], as well as multitask learning [S15] (details in Methods).

## S4. BOOSTING THE BASE REGRESSOR

The idea of improving a base regressor by examining the residuals originated in Tukey's work on reroughing [S16]. In this approach, we assume that an observed single-target can be decomposed into a sum of an underlying process that evolves smoothly, called the smooth term, and of an unsystematic noise component, called the noise term. In the context of the application, the smooth term in Eq. 2 is generated by the implemented approximation of the Hamiltonian model Eq. 1, and the noise term in Eq. 2 aims to capture generalizable physical effects missed by the classical control program (see Fig. 3).

The algorithmic paradigm of boosting originated from a question of Kearns and Valiant, about whether a weak learning algorithm that performs slightly better than random guessing, can be improved into an arbitrarily accurate strong learning algorithm, while working in the probably approximately correct (PAC) learning model [S17, S18]. In the affirmative, Schapire proposed the first provable polynomial-time boosting algorithm [S19], and Freund developed a more efficient boosting algorithm [S20]. Next, Freund and Schapire introduced AdaBoost [S21], which surmounted many of the practical difficulties of the earlier boosting algorithms. Then, Schapire et al. devised a way to incorporate prior knowledge into AdaBoost for single-target classification tasks, whereby prior knowledge is refined and not entirely overwhelmed by the process of learning from examples [S22]. The modification of the logistic loss function in AdaBoost arose in the development of spoken-dialogue systems at AT&T [S22].

The work of Friedman, Hastie, and Tibshirani linked the original formulation of AdaBoost [S21] with additive expansions [S23], which are typically fit with a backfitting algorithm or a greedy stagewise algorithm [S6, S24, S25]. Later, Breiman showed that boosting can be interpreted as a form of gradient descent in function space [S26]. Friedman extended this idea to the gradient boosting machine, which advantageously allows any choice of differentiable loss function [S27]. Simultaneously, Mason et al. developed an abstract characterization of boosting algorithms as gradient descent on empirical loss func-

tionals in an inner-product function space [S28]. Further connections with statistics were established in the work of Bühlmann and Hothorn [S29]. In practice, these algorithmic techniques are usually implemented in scikit-learn [S8], XGBoost [S30], or LightGBM [S31].

### A. Gradient boosting

In the main body, we devised a way to incorporate prior knowledge into the gradient boosting machine Alg. 1 (details in Methods). In this section, we expand upon this discussion by deriving a generic version of boosting for the $j$th single-target regression subtask, which follows from similar work in the original gradient boosting machine for single-target regression tasks [S6, S26–S29]. Let us begin by considering the $j$th empirical loss functional

$$\mathcal{L}_j(h_j) = \sum_{i=1}^{m_{train}} \ell_j(Y_j^{(i)}, h_j(X^{(i)})), \quad (S4)$$

where $\mathcal{L}_j$ is a function of the $j$th single-target regressor $h_j$. In this abstract characterization of boosting, the goal of a boosting algorithm is to minimize Eq. S4. As $\mathcal{L}_j$ is a functional, this minimization problem can be viewed as numerical optimization in function space

$$\mathbf{h_j^*} = \operatorname*{argmin}_{\mathbf{h_j}} \mathcal{L}_j(\mathbf{h_j}), \quad (S5)$$

where the parameter vector $\mathbf{h_j} \in \mathbb{R}^{m_{train}}$ components are the values of the $j$th single-target approximating regressor $h_j(X^{(i)})$ at each of the $m_{train}$ examples in the training data. Namely,

$$\mathbf{h_j} = \begin{pmatrix} h_j(X^{(1)}) \\ h_j(X^{(2)}) \\ \vdots \\ h_j(X^{(m_{train})}) \end{pmatrix}. \quad (S6)$$

Typically, numerical optimization procedures solve Eq. S5 by making an initial guess $\mathbf{h_{j,0}} = \mathbf{b_{j,0}} \in \mathbb{R}^{m_{train}}$, then iteratively updating each successive parameter vector $\mathbf{h_{j,k}}$ based on the current parameter vector $\mathbf{h_{j,k-1}}$, where we denote the number of iterations by $K_j$ and the subscript in $\mathbf{h_{j,k}}$ denotes the $j$th single target regression subtask and the $k$th iteration, respectively. Namely, we posit the solution of Eq. S5 as an additive expansion of parameter vectors

$$\mathbf{h_{j,K_j}} = \sum_{k=0}^{K_j} \mathbf{b_{j,k}}, \ \mathbf{b_{j,k}} \in \mathbb{R}^{m_{train}}. \quad (S7)$$

Here, we choose the stagewise fist-order functional steepest descent as the numerical optimization procedure. We have that each parameter vector is given by

$$\mathbf{b_{j,k}} = -\rho_{j,k}\mathbf{g_{j,k}}, \quad (S8)$$

where $\rho_{j,k} \in \mathbb{R}$ is the step length and $\mathbf{g_{j,k}} \in \mathbb{R}^{m_{train}}$ is the gradient of the empirical risk functional Eq. S4 evaluated at $\mathbf{h_j} = \mathbf{h_{j,k-1}}$; see reference [S32] for a derivation with the second-order functional Newton-Raphson update. Next, we compute each component of the gradient

$$g_{j,k}^{(i)} = \left.\frac{\partial \ell_j(Y_j^{(i)}, h_j(X^{(i)}))}{\partial h_j(X^{(i)})}\right|_{h_j(X^{(i)})=h_{j,k-1}(X^{(i)})}, \quad (S9)$$

as well as the step length

$$\rho_{j,k} = \operatorname*{argmin}_{\rho} \mathcal{L}_j(\mathbf{h_{j,k-1}} - \rho\mathbf{g_{j,k}}). \quad (S10)$$

Then, we make the update

$$\mathbf{h_{j,k}} = \mathbf{h_{j,k-1}} - \rho_{j,k}\mathbf{g_{j,k}}, \quad (S11)$$

and we repeat this process iteratively. We refer to this process as functional gradient descent [S6, S26–S29, S32].

In its current form, functional gradient descent does not address the generalization objective of a machine learning algorithm, as the gradient is only defined at a fixed set of $m_{train}$ examples and it cannot be generalized to other examples in the domain. In accord with the original gradient boosting machine [S6, S25, S27, S28], we resolve this dilemma by inducing a basis function $b(X;\theta)$, such as a shallow decision tree [S6, S27, S30, S31], at the $k$th iteration, which approximates the negative gradient signal. Hereby learning the parameter set $\theta_{j,k}$. We note that the parameter set $\theta_{j,k}$ would encode the split features, split locations, and the terminal node means of the individual trees for the $j$th single-target regression task at the $k$th iteration, if the choice of basis function were a shallow decision tree; see for example [S6, S27, S30, S31]. Next, we perform a one-dimensional line search

$$\alpha_{j,k} = \operatorname*{argmin}_{\alpha} \sum_{i=1}^{m_{train}} \ell_j(Y_j^{(i)}, h_{j,k-1}(X^{(i)})+\alpha b(X^{(i)};\theta_{j,k})), \quad (S12)$$

where $\alpha \in \mathbb{R}$; see reference [S29] for an argument about the possible omission of this step. Then, we sequentially append the induced basis function to the additive expansion

$$h_{j,k}(X) = h_{j,k-1}(X) + \alpha_{j,k} b(X;\theta_{j,k}). \quad (S13)$$

This leads us to the generic version of the $j$th single-target gradient boosting algorithm Alg. 1.

## S5. IMPLEMENTATION

We use the machine learning approach described in the main body, as well as the scikit-physlearn repository developed by Alex Wozniakowski. The Python based repository will be made publicly available at the cited GitHub link [S33]. For each single-target regression subtask, the training process takes a few seconds on a standard laptop,

and the hyperparameters can be accessed in the repository.

In obtaining the test error results in Fig. 2 and in Fig. S4, we evaluate all operations encompassing training data through nested cross-validation, which is executed in $GridSearchCV$ with parameters cv = 5 and scoring = "neg_mean_absolute_error" [S8]. For each single-target regression subtask, we independently pre-process the examples in Eq. 11 with a normal distribution quantile transformer [S8], and the transformer is handled by a modified pipeline object, which inherits from the pipeline object in reference [S8]. The modified pipeline object induces single-target regressors with Alg. 2, which takes as input the transformed training examples Eq. 11, number of iterations $K_j = 1$, squared error and absolute error single-target loss functions $\{\ell_j, \tilde{\ell}_j\}$, respectively, stacking regressor [S8, S11, S12] basis function $b$, number of cross-validation folds $k = 5$, and 5-fold cross-validation single-target absolute error loss function. In the stacking regressor [S8] with parameters cv = 5 and passthrough = True, the first stacking layer consists of a gradient boosted decision tree [S31] and a fully-connected neural network [S8], and the second stacking layer consists of a fully-connected neural network [S8]. The gradient boosted decision tree [S31] optimizes absolute error Eq. 5 and the neural networks optimize squared error Eq. S1. Moreover, we restrict the neural networks to one hidden layer in order to avoid overfitting, the activation function is the hyperbolic tangent function, and the optimization algorithm is the limited-memory variant of the Broyden-Fletcher-Goldfarb-Shanno algorithm.

As noted in the main body, Alg. 1 wins the model selection step in Alg. 2, so Alg. 2 makes a call to Alg. 1, which takes as input transformed training examples Eq. 11, number of iterations $K_j = 1$, squared error and absolute error single-target loss functions $\{\ell_j, \tilde{\ell}_j\}$, respectively, stacking regressor [S8, S11, S12] basis function $b$. In line (a) squared error $\ell_j$ is used. In line (b) the parameter sets are induced with the stacking regressor [S8]. In line (c), we appended an L$^1$ regularization term to the optimization problem

$$\underset{\alpha}{\mathrm{argmin}} \sum_{i=1}^{m_{train}} \ell_j(Y_j^{(i)}, h_{j,k-1}(X^{(i)}) + \alpha b(X^{(i)}; \theta_{j,k})) + \lambda |\alpha|.$$

We solve the optimization problem with the Nelder-Mead method, where $\lambda = 0.1$. In future applications, several modifications of Alg. 1 may be of interest: inclusion of other regularization terms, early stopping, out-of-bag-error estimates, or sampling techniques for variance reduction; see references [S6, S8, S25, S27].

In plotting the augmented learning curves in Fig. 4 and in Fig. S5, we modify the source code in reference [S8] to use the same withheld folds for the cross-validation error and the incumbent error. The training sizes in the plots from left to right are 23, 25, 27, 29, 31, 32, 34, 36, 38, 40, 42, 43, 45, 47, 49, 51, 52, 54, 56, 58, 60, 62, 63, 65, 67, 69, 71, 73, 74, 76, 78, 80, 82, 84, 85, 87, 89, 91, 93, 95, respectively. In plotting the summary plots in Fig. 5 and in Fig. S8, we utilize the SHAP framework [S34]. For each summary plot, we induce a fully-connected neural network [S8]. We restrict the neural network to one hidden layer, the activation function is the rectified linear unit, and the optimization algorithm is the limited-memory variant of the Broyden-Fletcher-Goldfarb-Shanno algorithm. We compute the SHAP values with Kernel SHAP [S34].

## S6. AUGMENTED LEARNING CURVES

In the main body, we illustrated the inbuilt model selection step in Alg. 2 with an augmented learning curve for the single-target regression subtask $Y_3$ with training sizes varying between 23 to 95 ordered pairs in Fig. 4. Here, we illustrate the inbuilt model selection step in Alg. 2 with augmented learning curves for the single-target regression subtasks $Y_1$ (top), $Y_2$ (top middle), $Y_4$ (bottom middle), and $Y_5$ (bottom) with training sizes varying between 23 to 95 ordered pairs in Fig. S5. For the single-target regression subtasks $Y_1$ (top), $Y_2$ (top middle), and $Y_5$ (bottom), the candidate always performs better than the incumbent. For the single-target regression subtask $Y_4$ (bottom middle), when there are less than 60 ordered pairs, the incumbent usually performs better, whereas the candidate always outperforms the incumbent with 60, or more, ordered pairs. These augmented learning curves highlight the importance of the inbuilt model selection step in Alg. 2, as the incorporation of prior knowledge into Alg. 1 does not always improve performance over the base regressor.

## S7. UTILITY OF THE DATA PREPROCESSOR

In this section, we describe the tabula rasa learning of the multi-targets [S15, S35–S39], which enables our utility study of the classical control program [S2, S3, S5] as a data preprocessor for the downstream boosting algorithm in the learning framework (see Fig. 1). Moreover, we show the summary plots [S34] for the single-target regression subtasks $Y_j$, where $j \in \{2, 3, 4, 5\}$.

In this setting, let $\mathcal{X} \subseteq \mathbb{R}^9$ be the domain of qubit and coupler biases [S1–S3, S5] in an instance of the many-body Ramsey spectroscopy technique [S2]. Let $\mathcal{Y} \subseteq \mathbb{R}^5$ be the target space of multi-target observations, as in the main body. We are given a finite sequence of labeled examples Eq. 3, where we regard a collection of 5 qubit and 4 coupler bias features as an example, and the associated multi-target observation as the label. Further, we retain the same data split as in the main body, so there are $m_{train} = 95$ training examples and $m_{test} = 41$ test examples.

We represent the training data with a $95 \times 14$ design

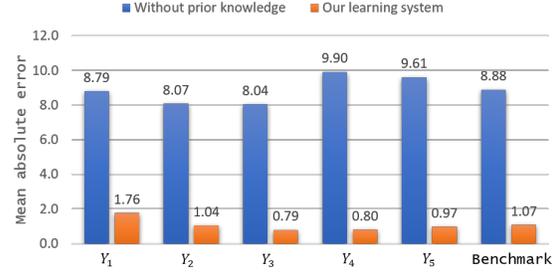

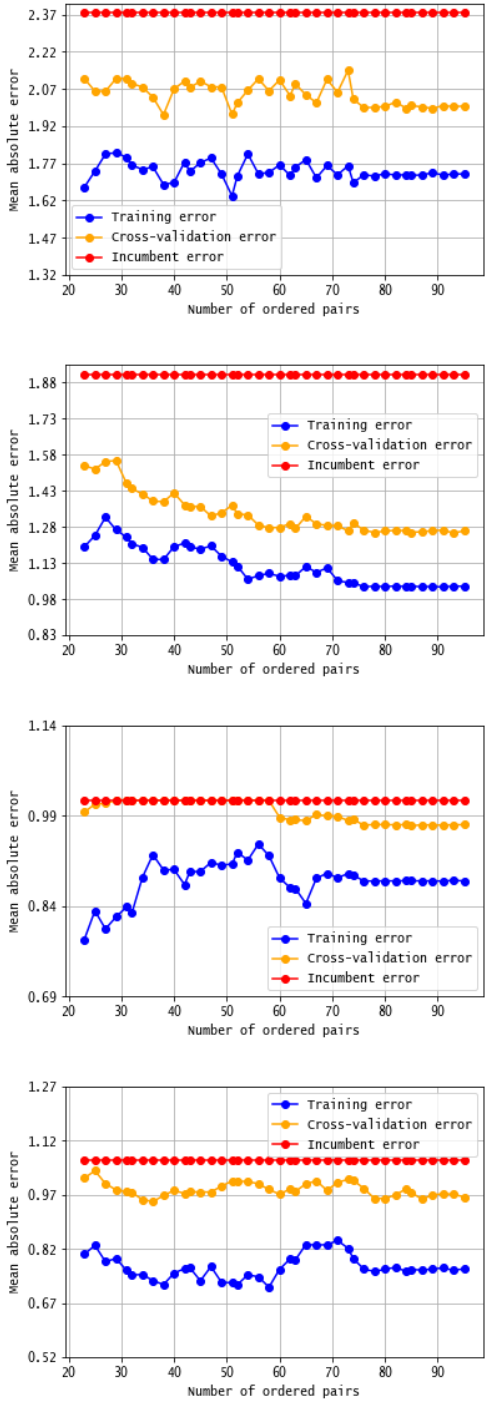

FIG. S5. Augmented learning curves for the single-target regression subtasks $Y_1$, $Y_2$, $Y_4$, and $Y_5$, from top-to-bottom respectively. We show the training, cross-validation, and incumbent errors for varying amounts of training examples in blue, orange, and red, respectively.

FIG. S6. Using the absolute error loss function as the criterion, we compare the best performing single-target regressors without prior knowledge against our learning system with prior knowledge in blue and orange, respectively. Disregarding the prior knowledge degrades performance by over 729% on the benchmark task.

matrix

$$\begin{pmatrix} \underline{\quad X^{(1)} \quad} & \underline{\quad Y^{(1)} \quad} \\ \underline{\quad X^{(2)} \quad} & \underline{\quad Y^{(2)} \quad} \\ \vdots & \vdots \\ \underline{\quad X^{(m_{train})} \quad} & \underline{\quad Y^{(m_{train})} \quad} \end{pmatrix}. \tag{S14}$$

Next, we slice Eq. S14 into single-target training data Eq. 11 for each single-target regression subtask $Y_j$, where $j \in \{1, 2, \ldots, 5\}$ and the shape of each matrix is $95 \times 10$. We employ model selection to choose the best performing single-target regressor for each single-target regression subtask, while restricting the search to a single complex model (details in Methods). We find a fully-connected neural network [S8] with a single-hidden layer as the best performing single-target regressor for each single-target regression subtask. After inducing each fully-connected neural network on Eq. 11, we evaluate the test error using the absolute error loss function in Fig. S6. Disregarding the prior knowledge degrades performance by over 729% on the benchmark task. As we have already generated the test example predictions, we utilize the squared error loss function to measure the prediction performance in Fig. S7. Disregarding the prior knowledge degrades performance by over 5426% on the multi-target regression task.

In the main body, we attained an overview of the most important qubit and coupler bias features for the single-target regression subtask $Y_1$ in Fig. 5. Here, we show the SHAP summary plots for the remaining single-target regression subtasks in Fig. S8, where $Y_2$ (top), $Y_3$ (top middle), $Y_4$ (top bottom), and $Y_5$ (bottom).

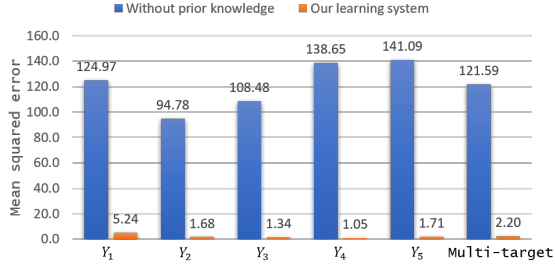

FIG. S7. Using the squared error loss function as the criterion, we compare the best performing single-target regressors without prior knowledge against our learning system with prior knowledge in blue and orange, respectively. Disregarding the prior knowledge degrades performance by over 5426% on the multi-target regression task.

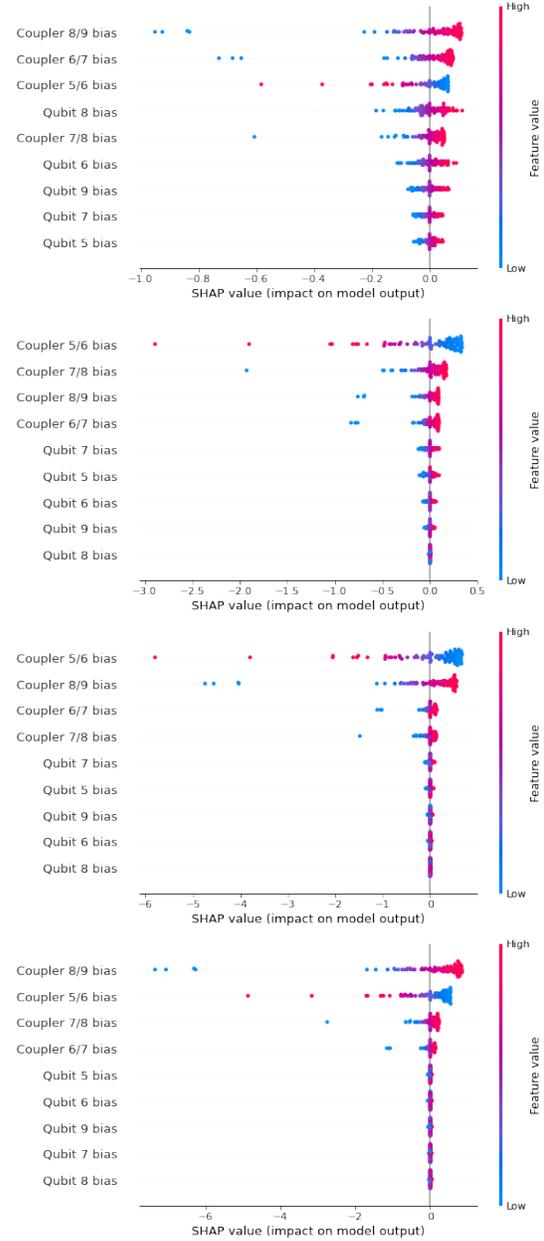

FIG. S8. SHAP summary plots [S34] for single-targets $Y_2$, $Y_3$, $Y_4$, and $Y_5$, from top-to-bottom respectively. In each plot, we use the collection of 9 qubit and coupler bias features as an example, and we induce a fully connected neural network [S8] to predict $Y_j$, where $j \in \{2, 3, 4, 5\}$. The horizontal axis is centered at the average training example prediction, and the vertical axis ascendingly orders the features according to their importance. Each point is a SHAP value for a particular example, the coloring represents the bias value, and overlapping points are randomly jittered along the vertical axis to avoid collisions [S34, S40].

[S1] Yu Chen et al. Qubit architecture with high coherence and fast tunable coupling. *Physical Review Letters*, 113:220502, 2014.

[S2] Pedram Roushan et al. Spectroscopic signatures of localization with interacting photons in superconducting qubits. *Science*, 358:1175–1179, 2017.

[S3] Charles Neill et al. A blueprint for demonstrating quantum supremacy with superconducting qubits. *Science*, 360:195–199, 2018.

[S4] Julian Kelly et al. Physical qubit calibration on a directed acyclic graph. arXiv:1803.03226, 2018.

[S5] Benjamin Chiaro et al. Growth and preservation of entanglement in a many-body localized system. arXiv:1910.06024, 2019.

[S6] Trevor Hastie, Robert Tibshirani, and Jerome Friedman. *The Elements of Statistical Learning*. Springer, 2009.

[S7] Gavin Cawley and Nicola Talbot. On over-fitting in model selection and subsequent selection bias in performance evaluation. *Journal of Machine Learning Research*, 11:2079–2107, 2010.

[S8] Fabian Pedregosa et al. Scikit-learn: machine learning in python. *Journal of Machine Learning Research*, 12:2825–2830, 2011.

[S9] Hanen Borchani et al. A survey on multi-output regression. *Wiley Interdisciplinary Reviews: Data Mining and Knowledge Discovery*, 2015.

[S10] Willem Waegeman, Krzysztof Dembczyński, and Eyke Hüllermeier. Multi-target prediction: A unifying view on problems and methods. *Data Mining and Knowledge Discovery*, 33:293–324, 2019.

[S11] David Wolpert. Stacked generalization. *Neural Networks*, 5:241–259, 1992.

[S12] Leo Breiman. Stacked regressions. *Machine Learning*, 24:49–64, 1996.

[S13] Leo Breiman and Jerome Friedman. Predicting multivariate responses in multiple linear regression. *Royal Statistical Society Series B*, 59:3–54, 1997.

[S14] Dumitru Erhan et al. Why does unsupervised pre-training help deep learning? *Journal of Machine Learning Research*, 11:625–660, 2010.

[S15] Rich Caruana. Multitask learning. *Machine Learning*, 28:41–75, 1997.

[S16] John Tukey. *Exploratory Data Analysis*. Addison-Wesley, 1977.

[S17] Michael Kearns and Leslie Valiant. Learning boolean formulae or finite automata is as hard as factoring. Technical Report TR-14-88, Harvard University Aiken Computation Laboratory, 1988.

[S18] Michael Kearns and Leslie Valiant. Cryptographic limitations on learning boolean formulae and finite automata. *Journal of the Association for Computing Machinery*, 41:67–95, 1994.

[S19] Robert Schapire. The strength of weak learnability. *Machine Learning*, 5:197–227, 1990.

[S20] Yoav Freund. Boosting a weak learning algorithm by majority. *Information and Computation*, 121:256–285, 1995.

[S21] Yoav Freund and Robert Schapire. A decision-theoretic generalization of on-line learning and an application to boosting. *Journal of Computer and System Sciences*, 55:119–139, 1997.

[S22] Robert Schapire et al. Incorporating prior knowledge into boosting. In *ICML'02: Proceedings of the Nineteenth International Conference on Machine Learning*, pages 538–545, 2002.

[S23] Trevor Hastie and Robert Tibshirani. *Generalized Additive Models*. Chapman and Hall, London, 1990.

[S24] Jerome Friedman, Trevor Hastie, and Robert Tibshirani. Additive logistic regression: A statistical view of boosting. *The Annals of Statistics*, 28(2):337–407, 2000.

[S25] Jerome Friedman and Bogdan Popescu. Importance sampled learning ensembles. Technical report, Stanford University, Department of Statistics, 2003.

[S26] Leo Breiman. Arcing the edge. Technical report, Stanford University, Department of Statistics, 1997.

[S27] Jerome Friedman. Greedy function approximation: A gradient boosting machine. *Annals of Statistics*, 29(5):1189–1232, 2001.

[S28] Llew Mason et al. Boosting algorithms as gradient descent. In *NIPS: Proceedings of the 12th International Conference on Neural Information Processing*, pages 512—518, 1999.

[S29] Peter Bühlmann and Torsten Hothorn. Boosting algorithms: Regularization, prediction and model fitting. *Statistical Science*, 22(4):477–505, 2007.

[S30] Tianqi Chen and Carlos Guestrin. XGBoost: a scalable tree boosting system. In *KDD'16: Proceedings of the 22nd ACM SIGKDD International Conference on Knowledge Discovery and Data Mining*, pages 785–794, 2016.

[S31] Guolin ke et al. LightGBM: a highly efficient gradient boosting decision tree. In *Advances in Neural Information Processing Systems 30*, pages 3149–3157. Curran Associates, Inc., 2017.

[S32] Fabio Sigrist. Gradient and newton boosting for classification and regression. arXiv:1808.03064, 2018.

[S33] Alex Wozniakowski. Scikit-physlearn. github.com/a-wozniakowski/scikit-physlearn, 2020.

[S34] Scott Lundberg and Su-In Lee. A unified approach to interpreting model predictions. In *Advances in Neural Information Processing Systems 30*, pages 4765–4774. Curran Associates, Inc., 2017.

[S35] David Hume. *A Treatise of Human Nature*. Clarendon Press, 1739.

[S36] David Haussler. Quantifying inductive bias: AI learning algorithms and Valiant's learning framework. *Artificial Intelligence*, 36:177–221, 1988.

[S37] Tom Mitchell. The need for biases in learning generalisation. In Jude Shavlik and Thomas Dietterich, editors, *Readings in Machine Learning*. Morgan Kaufmann, 1991.

[S38] David Wolpert and William Macready. No free lunch theorems for optimization. *IEEE Transactions on Evolutionary Computation*, 1(1), 1997.

[S39] Jonathan Baxter. A model of inductive bias learning. *Journal of Artificial Intelligence Research*, 12:149–198, 2000.

[S40] Christoph Molnar. Interpretable machine learning: a guide for making black box models explainable. christophm.github.io/interpretable-ml-book, 2020.



\end{document}